\documentclass[aps,rmp,reprint,amsmath,amssymb,graphicx,longbibliography,floatfix]{revtex4-1}
\usepackage{amssymb,amsfonts,amsmath} 
\usepackage{graphicx}
\usepackage{color}
\usepackage{xcolor}
\usepackage{graphics}
\usepackage{bm}
\usepackage{booktabs}
\usepackage[breaklinks=true]{hyperref}
\hypersetup{bookmarksnumbered, pdfpagemode=UseOutlines, 
pdfdisplaydoctitle, 
colorlinks=true, citecolor=blue, filecolor=blue, linkcolor=blue, urlcolor=blue}

\definecolor{colorp1}{rgb}{0.502,0.7686,0.8706}
\definecolor{color0}{rgb}{0.2275,0.4627,0.6353}
\definecolor{colorm1}{rgb}{0.6039,0.1804,0.1569}

\definecolor{colormc}{rgb}{0.5,0.2,0.5}

\newcommand*{\affilbrown}{Department of Physics, Brown University, Providence, Rhode Island 02912, USA}
\newcommand*{\affilaachen}{Institut f\"ur Theorie der Statistischen Physik, RWTH Aachen University and JARA-Fundamentals of Future Information Technology, 52056 Aachen, Germany}

\newcommand*{\affilmpsd}{Max Planck Institute for the Structure and Dynamics of Matter, Center for Free-Electron Laser Science (CFEL), Luruper Chaussee 149, 22761 Hamburg, Germany}
\newcommand*{\affilpenn}{Department of Physics and Astronomy, University of Pennsylvania, Philadelphia, PA 19104, USA}
\newcommand*{\affilpsi}{Laboratory for Micro and Nanotechnology, Paul Scherrer Institut, Forschungsstrasse 111, CH-5232 Villigen PSI, Switzerland}

\begin{document}

\title{{\it Colloquium:} Nonthermal pathways to ultrafast control in quantum materials}

\author{Alberto de la Torre}
\email{alberto\_de\_la\_torre\_duran@brown.edu}
\affiliation{\affilbrown}
\author{Dante M.~Kennes}
\email{dante.kennes@rwth-aachen.de}
\affiliation{\affilaachen}
\affiliation{\affilmpsd}
\author{Martin Claassen}
\email{claassen@sas.upenn.edu}
\affiliation{\affilpenn}
\author{Simon Gerber}
\email{simon.gerber@psi.ch}
\affiliation{\affilpsi}
\author{James W.~McIver}
\email{james.mciver@mpsd.mpg.de}
\affiliation{\affilmpsd}
\author{Michael A.~Sentef}
\email{michael.sentef@mpsd.mpg.de}
\affiliation{\affilmpsd}

\date{\today{}}

\begin{abstract}
We review recent progress in utilizing ultrafast light-matter interaction to control the macroscopic properties of quantum materials. Particular emphasis is placed on photoinduced phenomena that do not result from ultrafast heating effects but rather emerge from microscopic processes that are inherently nonthermal in nature. Many of these processes can be described as transient modifications to the free energy landscape resulting from the redistribution of quasiparticle populations, the dynamical modification of coupling strengths and the resonant driving of the crystal lattice. Other pathways result from the coherent dressing of a material's quantum states by the light field. We discuss a selection of recently discovered effects leveraging these mechanisms, as well as the technological advances that led to their discovery.  A road map for how the field can harness these nonthermal pathways to create new functionalities is presented.
\end{abstract}

%\pacs{81.05.Uw,68.37.-d,73.20-r}

\maketitle
\tableofcontents

\section{Introduction}
\label{intro}

\begin{figure*}[t]
  \begin{center}
    \includegraphics[width=\textwidth,clip,page=2]{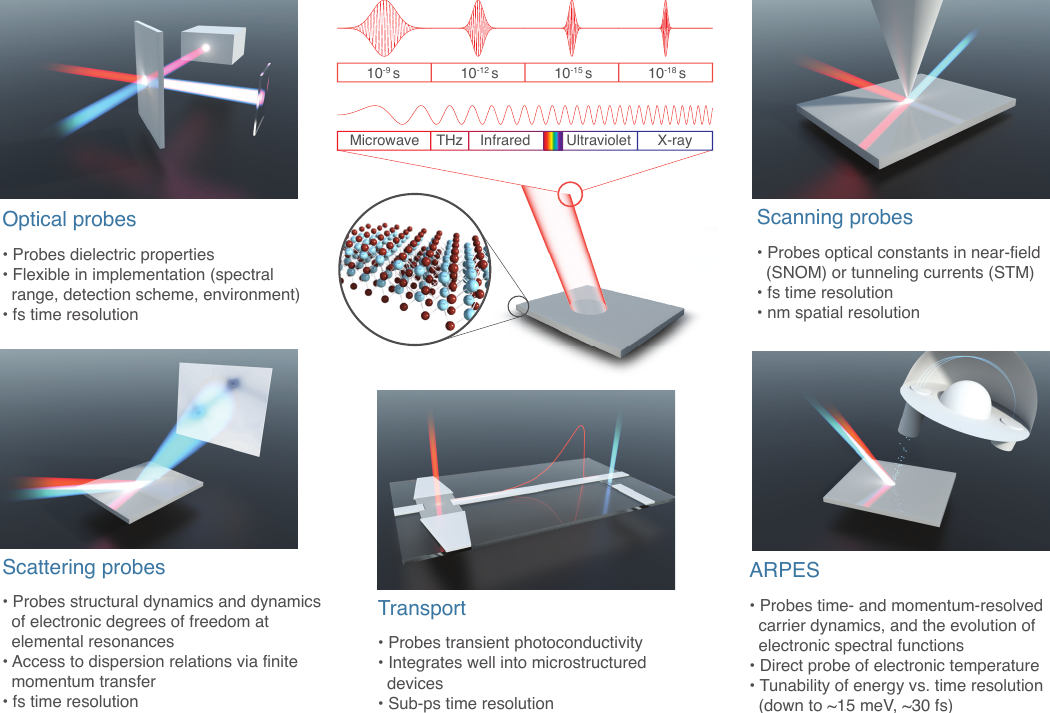}
  \caption{%{\bf 
  Overview of the main experimental tools available to investigate ultrafast phenomena in quantum materials. The upper central panel illustrates the broad spectral and temporal ranges from which tailored short laser pulses can be created to photoexcite (‘pump’) materials out of equilibrium. The surrounding panels show categories of ultrafast time-resolved probes and their capabilities.} 
  \label{fig:temp1}
  \end{center}
\end{figure*}

Quantum materials host a wide range of many-body and topological phenomena that both challenge our physical understanding of solids and offer possibilities for next-generation technologies. From unconventional superconductivity to topologically protected edge modes, the remarkable physics in quantum materials emerges from complex interactions between spin, charge, lattice, and orbital degrees of freedom \cite{Keimer2017} and the geometric and topological aspects of their wavefunctions \cite{wang_topological_2017,narang_topology_2021}. Many materials host multiple quantum phases that can be independently accessed by application of external perturbations such as electromagnetic fields, pressure, strain, or chemical doping, making these systems ideal platforms for future technological applications \cite{Tokura2017}. Moreover, the search for new ways to create and control their macroscopic properties continues to improve our fundamental understanding of the interactions between the underlying degrees of freedom, which in turn may lead to new functionalities \cite{basov_towards_2017,roadmap_QM_2020,alexandradinata_future_2020}.

One promising route to controlling and understanding quantum materials is ultrafast light-matter interaction. This can induce long-lived nonequilibrium states with functionally-relevant properties that cannot be realized in thermal equilibrium. Since the pioneering work of \citealt{Koshihara90} on photoinduced phase transitions, ultrafast optical experiments have been used to investigate a variety of nonequilibrium effects in quantum materials. Breakthrough results in the past decade include the surgical decoupling of microscopic degrees of freedom in iron-based superconductors \cite{Gerber17}, ultrafast switching into hidden phases in transition metal dichalcogenides \cite{stojchevska_ultrafast_2014}, dynamically controlled microscopic interactions in correlated transition metal oxides \cite{forst_nonlinear_2011}, possible light induced superconductivity in organic compounds \cite{mitrano_possible_2016}, and photon dressed topological states in topological insulators \cite{GedikARPES1} and graphene \cite{McIver}. %These findings have been enabled by the development of a new generation of experimental and theoretical tools to probe and understand the ultrafast processes in these systems ((Fig.~\ref{fig:temp1}) and Appendix.~\ref{sec:appendix}).

The mechanisms behind many of the most intriguing nonequilibrium phenomena go beyond the simple melting of thermal states by laser-induced heating. Instead, they rely on distinct \textit{nonthermal pathways}, such as transient modifications to the free energy landscape or photon dressing effects, where quasi-thermal descriptions based on effective temperatures are insufficient or impossible. 

Here, we survey recent efforts in applying \textit{nonthermality as a resource} to exert control over quantum materials on ultrashort timescales in a flexible and reversible manner. We highlight a selection of phenomena that emerge \textit{after} pulsed laser excitation, including nonthermal switching mechanisms, out-of-equilibrium critical behavior, and the nonlinear control of microscopic couplings and phononics (Sec.~\ref {sec:emergent}).   
Another class of nonthermal phenomena are those that occur \textit{during} the duration of the laser pulse, where the microscopic degrees of freedom are strongly coupled to the amplitude, frequency and polarization of the light field, in some cases forming dressed states that can be described by Floquet theory. %These dressed states allow for a parametric modulation of system parameters, which are discussed in the context of nonlinear phononics. 
%For long pulse durations and thus quasi-periodic situations, photo-dressed states can be described using Floquet theory. 
%While well established in atomic systems, photon-dressing effects in quantum materials can lead to fundamentally new states of matter. 
We review recent realizations of Floquet engineering protocols in quantum materials and discuss new theoretical directions for future experiments (Sec.~\ref{sec:dressed}). 

%In Appendix~\ref{sec:appendix}, we introduce some of the most widely used ultrafast techniques together with recent works that showcase their capabilities. 
%These findings have been enabled by the development of a new generation of experimental and theoretical tools to probe and understand ultrafast processes in quantum matter. 
Recent progress in the field is due in part to the development of improved time-resolved experimental techniques. State-of-the-art light sources can now reliably generate intense optical pulses at wavelengths spanning from the THz to the extreme ultraviolet with a wide range of available pulse durations and repetition rates \cite{Reimann2007,Steinmeyer1507,Bartels376,BLiu,Cerullo2003}. These advances have enabled a surgical approach to photoexciting (‘pumping’) quantum materials in pump-probe experiments, for example by selectively accessing electronic transitions, directly coupling to collective modes, or intentionally avoiding such resonances entirely. This enhanced pump tunability, combined with a fleet of experimental probes targeting complementary macroscopic observables (Fig.~\ref{fig:temp1} and App.~\ref{sec:exp_advances}), has led to an improved understanding of nonequilibrium phenomena in solids and the nonthermal pathways leading to their creation. Moreover, these results have triggered complementary theoretical efforts towards modeling microscopic dynamics in photoexcited quantum materials (App.~\ref{sec:theory_advances}). 

A full-fledged review of the rapidly growing field of ultrafast quantum materials science is beyond the scope of this Colloquium. To account for the important early works, we refer to other reviews, including \citealt{Averitt2002}, \citealt{Basov11}, \citealt{orenstein_ultrafast_2012}, \citealt{Zhang14} and \citealt{giannetti_ultrafast_2016}, which provide extensive reports of ultrafast spectroscopy of strongly correlated electron systems. \citealt{Kirilyuk2010} wrote an in-depth review on ultrafast magneto-optical effects. \citealt{buzzi_probing_2018} reviewed ultrafast structural dynamics in solids probed by time-resolved X-ray scattering. Overviews of time-resolved inelastic X-ray scattering have been provided by \citealt{Cao_rev_19} and \citealt{Mitrano2020}. Further reviews on Floquet engineering can be found in \cite{bukov_review,Oka19Review,rudner_band_2020}.

\begin{figure*}[t]
  %\vspace{-10mm}
    \includegraphics[width=\textwidth,clip,page=3]{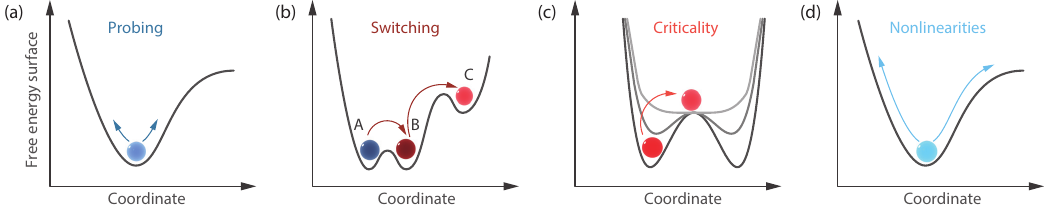}
    %\vspace{-8mm}
  \caption{%{\bf 
  Illustration of nonthermal pathways triggered by laser excitation. All panels show a potential or free energy landscape as a function of a system coordinate, e.g., an electronic order parameter or a lattice displacement. (a) Weak excitations around a stable minimum (ground state) permit probing collective modes and their mutual couplings. (b) With a short, more intense laser pulse, one can switch between degenerate ground states ($A$ and $B$), or drive the material to a nonthermal, metastable excited state $C$.
  (c) Nonequilibrium critical behavior can be induced by a short, even more intense excitation that transiently modifies the free energy landscape itself. (d) With a strong excitation, the system can also be driven into anharmonic regimes with deviations from parabolic behavior, which allows to probe nonlinear effects and change effective couplings in the material.} 
  \label{fig:section_iii}
\end{figure*}

\section{Emergent phenomena after pulsed laser excitation}
\label{sec:emergent}

We discuss here emergent phenomena in quantum materials occurring \textit{after} photoexcitation by an ultrafast laser pulse. %Comprehensive reviews of these effects can be found by \citealt{Basov11}, \citealt{orenstein_ultrafast_2012}, \citealt{Zhang14} and \citealt{giannetti_ultrafast_2016}. 
In particular, we focus on processes dominated by dynamics that cannot be described by a hot electronic subsystem which thermalizes back to equilibrium through heat exchange with the cold crystalline lattice (phonons), the so-called two-temperature models \cite{anisimov_1974,allen_theory_1987} or $n$-temperature generalizations \cite{koopmans_explaining_2010}. These descriptions rest on the assumption that the mutual couplings between degrees of freedom are not modified by the excitation \cite{allen_theory_1987,petek97,Bauer2015}. In quantum materials, this is not necessarily the case, and the optical excitation can result in modified or suppressed coupling constants \cite{ishioka_decoupling_2008,PhysRevB.91.045128}. More importantly, these simplified models neglect nonthermal effects resulting from ultrafast light-matter coupling. For example, in many correlated systems the notion of thermality, in the sense of a unique thermal density matrix and temperature, breaks down on ultrashort (femtosecond) to intermediate (sub-picosecond) timescales \cite{Kemper_general_principles_2018}. This departure from the thermal response makes the pulsed excitation of quantum materials a unique strategy for creating and controlling quantum phenomena.

%To better summarize this rapidly evolving and complex field, 
We depict in Fig.~\ref{fig:section_iii} different scenarios of transient phenomena in terms of a simplified free energy landscape as a function of growing excitation strength. In the weak-excitation regime, the optical drive can excite collective oscillations of the underlying order parameters (Fig.~\ref{fig:section_iii}(a), Sec.~\ref {sec:interplay}). By tracking these modes in the time domain one can disentangle microscopic degrees of freedom and their mutual couplings. In some materials, photoexcitation can drive the system into a different low-energy state or hidden states not accessible in thermal equilibrium (Fig.~\ref{fig:section_iii}(b), Sec.~\ref {sec:switching}). Under the right excitation conditions, ultrafast transient modifications of the free energy landscape can also induce nonequilibrium phase transitions and enable the study of nonthermal critical behavior (Fig.~\ref{fig:section_iii}(c), Sec.~\ref {sec:critical}). Finally, strong excitations can drive a system into nonlinear regimes, dynamically modifying the effective microscopic couplings governing its quantum many-body wavefunction, thus providing another strategy to investigate emergent phenomena with no equilibrium counterpart (Fig.~\ref{fig:section_iii}(d), Sec.~\ref {sec:nonlinearities}). 

\subsection{Weak-excitation regime: Collective modes and disentangling degrees of freedom}
\label{sec:interplay}

We consider a situation where the system under investigation is only weakly driven, i.e., it remains in a linear regime [Fig.~\ref{fig:section_iii}(a)]. The recovery towards equilibrium is sensitive to collective modes and reflects the characteristic timescales of the different degrees of freedom. Prominent examples of coherent modes in ordered states that have been probed using time-domain techniques are amplitude modes of charge-density wave materials \cite{Demsar99,Demsar2002,lee_nickelate_2012,Zong2019a,schmitt_transient_2008,Mihailovic19}, potential excitonic insulators \cite{Werdehausen2018}, and the Higgs amplitude \cite{matsunaga_higgs_2013,matsunaga_light-induced_2014,chu_phase-resolved_2020} and Leggett modes \cite{Giorgianni19}, as well as Josephson plasmons \cite{Laplace16} in superconductors. %Moreover, weak pump intensities reveal details about phase transitions \cite{Cavalleri05,Koshihara90,Beaud14} or microscopic couplings such as magnetic exchange interactions \cite{PhysRevLett.125.197203,Dean2016,Mazzone20}.

\begin{figure}[t]
  \begin{center}
    \includegraphics[width=0.5\textwidth,clip,page=4]{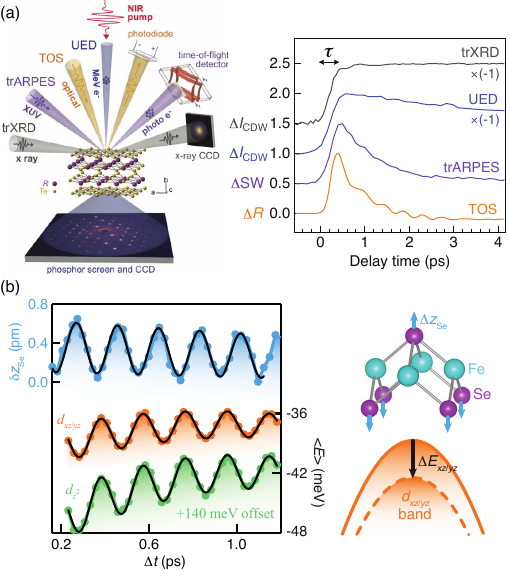}
  \caption{%{\bf 
  Multi-messenger probing of disentangled microscopic degrees of freedom. (a) A portfolio of ultrafast techniques has been used to investigate photoinduced melting of charge-density wave order in the materials (La/Tb)Te$_3$, including time-resolved X-ray diffraction (tr-XRD), time- and angle-resolved photoemission spectroscopy (tr-ARPES), transient optical spectroscopy (TOS) and ultrafast electron diffraction (UED). The signals obtained can be qualitatively and quantitatively different since distinct degrees of freedom are probed. Taken from \onlinecite{Zong2019a}. (b) Se atoms in FeSe are coherently displaced following photoexcitation by an ultrafast IR pulse, which can be directly measured by tr-XRD (blue, top). At the same time, corresponding energy-shifts of electron bands are revealed by tr-ARPES (orange, middle; green, bottom). The schematic on the right depicts the $A_{1g}$ phonon mode which induces the periodic modulation of the lattice and electronic bands. Taken from \onlinecite{Gerber17}.} 
  \label{fig:disentangle}
  \end{center}
\end{figure}

In the weak-excitation regime, charge, lattice, spin or orbital degrees of freedom of correlated electron systems can be disentangled in the time domain by appropriate choice of the experimental probe [Fig.~\ref{fig:disentangle}(a)]. 
Under particular circumstances these techniques can be brought together in a unified fashion, as demonstrated by \citealt{Gerber17} where time- and angle-resolved photoemission spectroscopy (tr-ARPES) and X-ray scattering were used to study the significance of a cooperative interplay among electron-electron and electron-phonon interactions in the iron-based superconductor parent compound FeSe. Figure~\ref{fig:disentangle}(b) depicts how both the lattice displacement and the electronic band structure lock into a coherent $A_{1g}$~phonon mode that is driven by an ultrafast infrared (IR) laser pulse. In turn, this THz frequency locking of the two degrees of freedom allows for a purely experimental, highly precise quantification of fundamental physical properties. For example, for FeSe it was used to quantify the orbitally-resolved electron-phonon deformation potential, which reveals the critical importance of electron correlations for the material's properties.

This example shows that such multi-messenger approaches not only allow inferring nonequilibrium characteristics, but also to extract equilibrium properties via the time domain \cite{Gerber17,Mitrano19}. A caveat concerns the difficulty in bringing together more than one ultrafast experiment in a unified fashion. Ideally, a multi-messenger experiment is conducted in one single setup, e.g. as used for simultaneous detection of electronic and structural \cite{Porer14} or multiple structural orders \cite{Morrison14,Kogar20,Zhou21}. This ensures that the pumping conditions, such as pump fluence, pump and probe beam profile, time resolution, penetration depth (effectively absorbed energy densities) and, notably, also the definition of \textit{time zero} (the instant when the pump pulse excites the sample) are guaranteed to be consistent. Such unified experimental setups will not only allow comparing the amplitude signals of the disentangled degrees of freedom, but also their phase relation and decay rates (dephasing times), paving the way towards coherent quantum control of \mbox{(dis-)entangled} degrees of freedom.

\subsection{Optical switching}
\label{sec:switching}

\begin{figure}[t]
    \includegraphics[width=\columnwidth,clip,page=5]{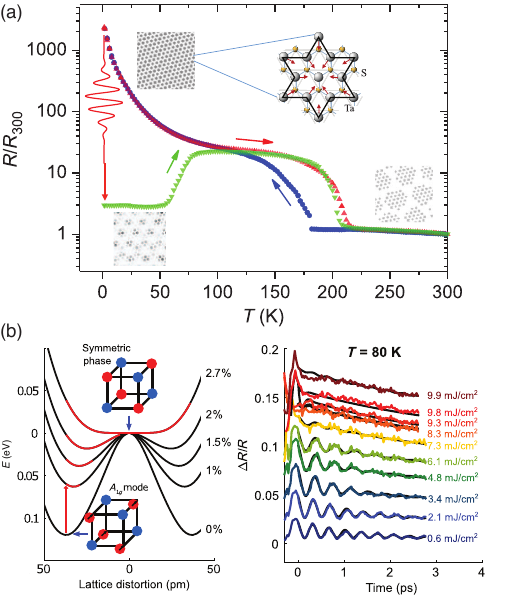}
  \caption{%{\bf 
  Ultrafast switching. (a) At equilibrium $1T$-TaS$_2$ goes through successive CDW transitions as a function of temperature, resulting in an insulating low-temperature state that is characterized by strong correlations (blue, circles; red, triangle up). Upon photoexcitation with a single 35-fs optical pulse, a long-lived metallic state emerges (green, triangle down). Adapted from \onlinecite{vaskivskyi_controlling_2015} © The Authors, some rights reserved; exclusive licensee AAAS. Distributed under a \href{http://creativecommons.org/licenses/by-nc/4.0/}{Creative Commons Attribution NonCommercial License 4.0 (CC BY-NC)}. (b) Photoexcitation of electrons in bismuth changes the local nuclear potential energy. For excitation densities above $2\%$ the amplitude of the initiated $A_{1g}$ mode drives a transition to a higher-symmetry phase. The $A_{1g}$~phonon mode, characteristic of the low symmetry phase, gets damped as the new phase is reached. Adapted from \onlinecite{Teitelbaum2018}.}
  \label{fig:switch}
\end{figure}

Going one step further away from equilibrium, a natural question to ask is whether nonthermal pathways can be found to photoinduce a phase transition, that is, optically \emph{switch} the system into parts of the free energy landscape that are not accessible in thermal equilibrium. %The first experimental demonstration of optical switching was provided by \citealt{Koshihara90} in organic crystals. 
Indeed, the key idea behind optically driven switching is similar to basic concepts of chemical reaction pathways, in which catalysts help the system overcome an activation energy barrier in order to trigger the intended reaction. 

We start with a pedagogical description within a free energy landscape framework, as sketched in Fig.~\ref{fig:section_iii}(b) \cite{Nasu_2001}. We consider a long-range ordered symmetry-broken ground state given by one of the global minima of the free energy landscape, indicated by $A$. In order to reach any of the local minima in its vicinity ($C$) the system has to overcome an associated energy cost that might be insurmountable in thermal equilibrium. However, excitation with an ultrashort laser pulse can lift the system over this energy barrier by nonadiabatic energy transfer into the electronic degrees of freedom and give rise to metastable phases \cite{Koshihara90,PIPT_Tokura}, which in some cases can persist indefinitely on experimental timescales, as first reported in manganites \cite{miyano_manganite_1997,Kiryukhin}. In $1T$-TaS$_2$ \cite{Fazekas_review}, the ultrafast transfer of energy into the electronic degrees of freedom, by either laser ($<35$~fs) \cite{stojchevska_ultrafast_2014,Perfetti06,Zong18,Danz21,Vodeb2019,Gerasimenko2019} or electrical ($\sim30$~ps) pulses \cite{vaskivskyi_controlling_2015,MihailovicCCM}, results in a long-lived amorphous-like metallic state, as illustrated in Fig.~\ref{fig:switch}(a), due to the formation of light-induced topological defects \cite{Haupt16,Laule17}. Another example of an ultrafast transition induced by nonthermal electronic configurations is the switching into long-lived high-symmetry crystalline phases in prototypical Peierls systems, as shown by \onlinecite{Teitelbaum2018,naseska2020firstorder,Huber14} [Fig.~\ref{fig:switch}(b)]. Ultrafast excitation of the electronic degrees of freedom in these materials results in a nonthermal distribution that thermalizes into the high-energy metastable state via dissipation of the surplus energy into a heat bath, e.g., phonons or the substrate \cite{Demsar_proceedings_2014}. The equilibrium state is then reinstated after a slow relaxation process or upon heating above a critical temperature \cite{Gerasimenko2019,vaskivskyi_controlling_2015}. For a comprehensive theoretical description of ultrafast optical switching in terms of a Ginzburg-Landau (GL) theory we point the reader to a the recent work by \citealt{SunMillis}.

Nonthermal metastable phases open new opportunities for controllable and reversible switching and thus promises potential for technological applications. This has motivated work in a variety of quantum material families. To exemplify the universality of optical switching, and to acknowledge their relevance within the ultrafast condensed matter field, we highlight here three research directions: \emph{(i)} light-induced insulator-to-metal transitions in nanofabricated \cite{liu_terahertz-field-induced_VO2_2012} and pristine samples of the correlated insulator VO$_2$ \cite{Becker_VO2,baum788,Otto450,Cavalleri_VO2_2001,Cavalleri_VO2_2004,Morrison14,Wall18,Wegkamp14}, \emph{(ii)} the control of the spin-orbit-lattice coupled ground state in manganites \cite{Tokura462} by either laser \cite{takubo_manganite_2005,ichikawa_transient_2011,li_femtosecond_2013,zhang_cooperative_2016,Teitelbaum2019,Beaud14,li_theory_2018} or electric field \cite{asamitsu_current_1997} pulses, and (\emph{iii}) the emergence of a long-lived state with superconducting-like properties at higher temperatures than at equilibrium upon light-induced suppression of the charge (stripe) order in the $1/8$-doped cuprates \cite{nicoletti_optically_2014,fausti_light-induced_2011,cremin_photoenhanced_2019} and in K$_3$C$_{60}$\cite{mitrano_possible_2016,Cantaluppi2018,budden_evidence_2021}.

These works have paved the way for moving the field towards developing new flexible, fast and reversible switching strategies, for example unimpeded by the healing of topological defects \cite{Mihailovic19,MihailovicCCM}. Here the field of multiferroic materials---in which magnetism is controlled by DC electric fields---has served as inspiration \cite{spaldin_multiferroics_2020} for ultrafast experiments that reported the engineering of ferroelectric states \cite{Stoica2019,Kubacka14}. Ultrafast switching into a metastable ferroelectric phase has been observed after coherently driving the quantum paraelectric phase of SrTiO$_3$ with strong THz fields \cite{Nova2019,Li2019}. This approach has also been used to demonstrate the ultrafast reversibility of the ferroelectric polarization in both directions, essential for future technological applications \cite{Qi09,mankowsky_ultrafast_2017}. Additionally, the exciting prospect of utilizing laser pulses to switch the topological state of a material has recently been demonstrated in THz-driven WTe$_2$ \cite{Sie2019}.

Finally, we discuss the key ideas behind symmetry-guided switching \cite{tsuji_pseudospin_resonance_2015,sentef_theory_2017,chou_twisting_pseudospins_2017,dehghani_dynamical_2017,Kennes:lightdwave,naga_eckstein_field_controlled_2019,ono_ishihara_2019,dehghani_optically_induced_topo_sc_2020} on the specific example of chiral superconductors \cite{claassen_universal_2019}, which constitutes a theoretically proposed but not yet experimentally verified ultrafast switching paradigm. Going back to Fig.~\ref{fig:section_iii}(b), this example corresponds to a free energy landscape with two minima $A$ and $B$ which represent the left- and right-handed chiral ground states of a topological superconductor. It was shown that a combination of linearly polarized followed by circularly polarized laser pulses can nudge the system from the right- to the left-handed chiral ground state. Importantly, the system can immediately be switched back equally fast by applying another sequence of pulses and simply changing the chirality of the circularly polarized laser pulse. The key ingredient besides laser polarization is the duration of the two pulses: the circular pulse has to be applied while the system has not yet recovered from the linear pulse and remains in a superposition of the two chiral states. Recently, it was shown theoretically how the same mechanism, but with spatially localized laser spots, could be used to write, move, or erase chiral domains in real space \cite{Tao2021}.

\subsection{Out-of-equilibrium critical behavior}
\label{sec:critical}

Ultrafast photoexcitation can induce an instantaneous change of the free energy landscape by suppressing the relevant order parameter [Fig.~\ref{fig:section_iii}(c)]. This phenomenology challenges our understanding of symmetry-breaking phase transitions. In equilibrium, the macroscopic time-dependent GL theory assumes an order parameter which varies `sufficiently' slowly in time. However, when the order parameter is suppressed on system-intrinsic timescales, it is unclear whether equilibrium concepts (universality classes, scaling) remain applicable, and how the order parameter fluctuations evolve after light excitation. Here we discuss out-of-equilibrium symmetry-breaking phase transitions both as a pathway towards accessing nonthermal effects but also as a platform to study fundamental questions about macroscopic properties associated with order parameter dynamics. The seminal works by \citealt{Hohenberg1977}, \citealt{PolkovnikovRMP11} and \citealt{Tauber_review} offer comprehensive reviews.

\begin{figure}[t]
  \begin{center}
    \includegraphics[width=0.45\textwidth,clip,page=7]{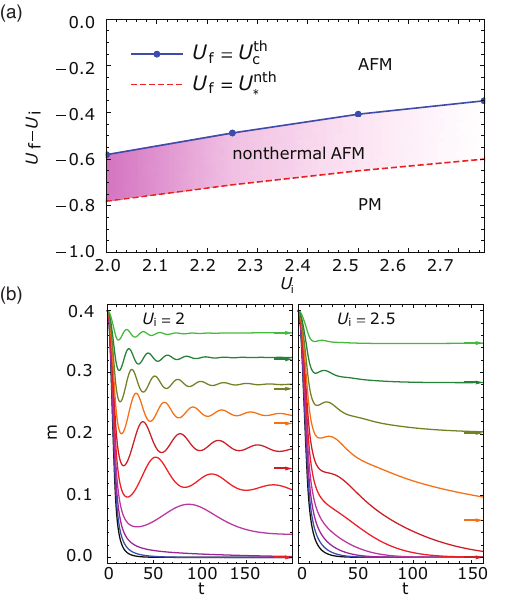}
  \caption{%{\bf 
  Nonthermal critical behavior. (a) Nonequilibrium phase diagram as a function of $U_{\rm i}$ and size of the quench $U_{\rm f} - U_{\rm i}$. A splitting of the critical point into thermal and nonthermal antiferromagnetic (AFM) critical points is observed. (b) Time evolution of the magnetization for a quench of the electron-electron interaction $U_{\rm i} \rightarrow U_{\rm f}$ in the antiferromagnetic phase of the Hubbard model. Each color represents a different value of $U_{\rm f}$ (left: $U_{\rm f}=1.0,1.1,\ldots,1.9$ and right: $U_{\rm f}=1.5,1.6,\ldots,2.4$ from bottom to top). Adapted from \citealt{Tsuji2013}.} 
  \label{fig:criticality}
  \end{center}
\end{figure}

An example of universal critical behavior in out-of-equilibrium phase transitions is the \textit{slowing down} of the recovery dynamics. Similarly to what occurs in thermal equilibrium, in the vicinity of a second-order transition, spatial fluctuations of the order parameter diverge, and the exponential rate with which the ordered phase is reinstated vanishes following a power law. This has been seen, for example, in the electronic recovery of charge order in perovskite manganites \cite{Beaud14} and  La$_{1/3}$Sr$_{2/3}$FeO$_3$ \cite{Zhu2018}, in the suppression dynamics of the charge density wave in LaTe$_3$ \cite{Zong2019a}, as well as in the spin and orbital dynamics in ${\mathrm{YVO}}_{3}$ and ${\mathrm{GdVO}}_{3}$\cite{Yusupov2010b}, and ${\mathrm{LaVO}}_{3}$ \cite{Lovinger20}. Divergent slow dynamics also characterize the ultrafast suppression of the nematic electronic order in FeSe \cite{Shimojima2019} and in iron pnictides \cite{Patz2014}. This seemingly universal behavior in out-of-equilibrium second-order phase transitions has been discussed in the time-dependent GL framework \cite{dolgirev_self-similar_2020}, by stochastic Langevin equations describing the relevant subsystems \cite{Sieberer2016}, and by non-equilibrium Green's functions \cite{Schuler18}.

Additionally, ultrafast phase transitions open the possibility to study the effects of nonadiabatic suppression. Near the phase transition the ordered and disordered phases are quasi-degenerate in energy. This allows the fluctuations of the order parameter to be arbitrarily slow. Within the Kibble-Zurek mechanism \cite{ZUREK1996,Kibble1976}, when the order parameter suppression happens on timescales faster than that of the fluctuations, coherence between order-parameter domains is lost and the critical behavior is replaced by recovery dynamics dominated by the healing of topological defects. This behavior is universal  \cite{Bray_phase_ordering_kinetics} and has been observed in the long-time recovery dynamics after suppression of the striped phase in cuprates \cite{Mitrano19} and of the charge order in LaTe$_3$ \cite{Zong2019}, TbTe$_3$ \cite{Yusupov2010,Mertelj_topodef_2013}, and 1$T$-TaS$_2$ \cite{Vogelgesang2018}. A different nonthermal scenario has been shown to emerge when two strongly coupled order parameters are quenched by a laser: In the striped nickelate La$_{1.75}$Sr$_{0.25}$NiO$_4$ an absence of topological defects was reported in conjunction with slow phase fluctuation recovery \cite{lee_nickelate_2012,chuang_nickelate_2013} and explained by a time-dependent GL theory \cite{kung_recovery_theory_2013}.   

We finally discuss predictions of nonthermal behavior in strongly correlated systems when driven through a phase transition \cite{Chiocchetta_2017}. For example, quenching of the electron-electron repulsion $U$ in the antiferromagnetic phase of the Hubbard model results in a suppression of the magnetization. Depending on the size of the quench it was shown that a nonzero order parameter and coherent amplitude mode oscillations survive even for quenched values of $U$ at which the system would be in the disordered phase in thermal equilibrium \cite{Werner2012,Tsuji2013,Sandri_nonequilibrium_2013,Balzer15,werner_murakami_nonthermal_excitonic_condensation_2020}. This gives rise to a prototypical out-of-equilibrium phase diagram shown in Fig.~\ref{fig:criticality}, where the suppression of the amplitude mode is associated with a secondary nonthermal critical point at $U_{\rm f}=U^{\rm nth}_{*}$. It is worth noting that such a transition is not expected within the GL picture, where at the critical point the amplitude mode oscillations vanish as the curvature of the free energy potential is suppressed and, consequently, the restoring force disappears \cite{Hohenberg1977}. 

%Thereby, out-of-equilibrium phase transitions open new opportunities to implement optical control by exploiting the different regimes defined by the intrinsic dynamics of each subsystem, where the dynamical transition is controlled via nonthermal populations of relevant collective modes, e.g., electrons, magnons, or phonons.

\subsection{Nonlinearities and dynamical couplings}
\label{sec:nonlinearities}
We now discuss cases where nonlinear regimes are reached through intense excitation and showcase two examples: (i) dynamical effective interactions due to modified electronic screening, in which the nonlinearity stems from electron-electron interactions, and (ii)~strongly driven crystal lattices, in which the nonlinearity is due phonon-phonon interactions or other phonon-related nonlinearities. We note that this classification is mainly for pedagogical reasons. In reality, in both cases the effective Hamiltonian is transiently modified, and it is not always straightforward to reveal the microscopic origin for this modification. The first direct experimental evidence of transiently changed electron-electron interactions was interpreted in the context of (ii) \cite{kaiser_optical_2014,singla_thz-frequency_2015}.

\subsubsection{Dynamical Hubbard \textit{U}}
\label{sec:dynamicalu}

\begin{figure}[htp!]
  \begin{center}
    \includegraphics[width=0.45\textwidth,clip,page=8]{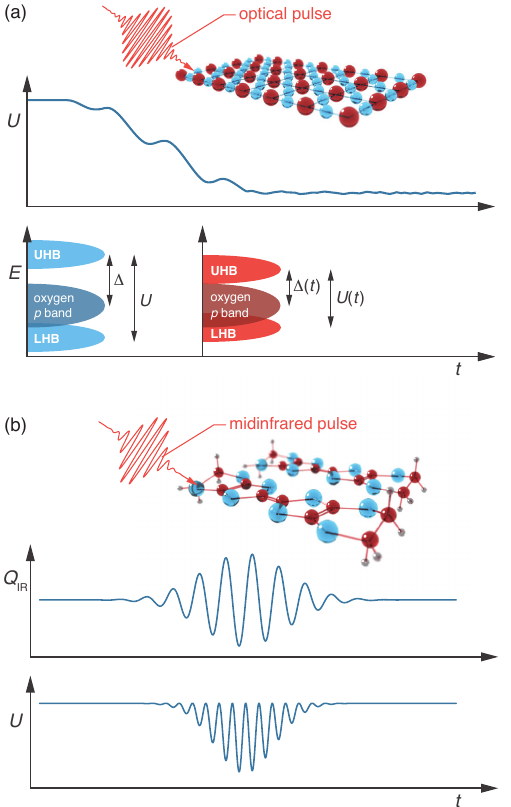}
  \caption{%{\bf 
  Dynamically screened Hubbard $U$ in laser-driven quantum materials. (a) Screening is typically enhanced when a material is excited by an optical laser pulse, which creates electronic excitations, resulting in a dynamically reduced time-dependent $U$ \cite{tancogne-dejean_ultrafast_2018,golez_multiband_2019}. In charge-transfer insulators this results in a narrowing of both the Mott-Hubbard gap between the upper and lower Hubbard bands (UHB and LHB, respectively), as well as the charge-transfer gap $\Delta$ \cite{PhysRevB.102.115106}. (b) For a mid-IR laser excitation, resonant with an IR-active phonon mode, the dynamical $U$ can parametrically depend on the phonon mode coordinate $Q_{\text{IR}}$ via a nonlinear electron-phonon interaction ($Q_{\text{IR}}^2 U$). %\cite{singla_thz-frequency_2015,buzzi_photomolecular_2020}.
  Whether $U$ increases or decreases on average depends on the sign of this nonlinearity.} 
  \label{fig:dynU}
  \end{center}
\end{figure}

The dynamical regime after ultrafast photoexcitation opens a pathway for changing the effective interactions in correlated materials. The paradigmatic example for such an effective interaction is the Hubbard $U$, which parametrizes the screened local part of the Coulomb integrals of strongly localized (typically $d$ or $f$) orbitals. Examples for applications of the Hubbard model are transition-metal oxides \cite{imada_metal-insulator_1998}. There, localized transition-metal ions are embedded between more delocalized ligand orbitals, which often are energetically separated from the $d$ orbitals. In such cases a downfolding procedure can be applied \cite{miyake_ab-initio_2009}, in which the low-energy effective model contains a Hubbard~$U$ that depends on the screening from electrons in delocalized environmental orbitals. The key idea of \textit{dynamical Hubbard $U$} is that the electronic excitations created after interaction with an ultrafast laser pulse can modify the screening environment, effectively changing $U$ within femtosecond timescales [Fig.~\ref{fig:dynU}(a)].

To explain the underlying mechanism, we discuss here the transient reduction of $U$ upon photoexcitation in the correlated antiferromagnetic charge-transfer insulator NiO predicted within time-dependent density functional theory plus $U$ (TDDFT+$U$) \cite{tancogne-dejean_ultrafast_2018}. The underlying physics can be understood as follows: 
The electronic structure of NiO is such that the localized correlated subspace of mainly $d$-orbital character and the rest of the system of mainly $p$-orbital character can be considered independently. Photoexcitation of electrons from the localized subspace into delocalized states results in additional screening. This naturally leads to a decrease of the effective $U$. It is instructive to think of the excitation as an induced time-dependent polarization of the localized electrons, affecting the dielectric properties of the delocalized electrons, which is expected for strong driving fields even for frequencies away from optical transitions. Similar results were obtained by \onlinecite{golez_multiband_2019,gillmeister_ultrafast_2020} using different methods. 
%Additionally, transient renormalization of $U$ was predicted to drive a transition from an antiferromagnetic insulator state into a topological semimetallic phase in magnetic pyrochlore iridates \cite{topp_all-optical_2018}.

Experimental evidence for a dynamically modified Hubbard $U$ by photoexcitation of electron-hole pairs was reported in the transition-metal dichalcogenide MoTe$_2$ \cite{beaulieu_ultrafast_2021}. 
In this material, the uncorrelated band structure has two electron-like bands crossing the Fermi level $E_{\rm F}$ at the edge of the Brillouin zone. However, this is not observed in photoemission. Instead, as explained by DFT+$U$, the Hubbard $U$ pushes the respective states above $E_{\rm F}$. This effect can be partially undone on ultrafast timescales, as demonstrated by tr-ARPES. $\sim200$~fs after photoexcitation, an ultrafast Lifshitz transition was observed, at which the Fermi surface topology changed because the electron-like bands were pushed below $E_{\rm F}$ triggered by dynamical reduction of $U$.

A different route towards dynamical modulation of Hubbard $U$ is the coherent driving of the crystal lattice. Dynamical $U$ through phonon driving was first demonstrated in a prototypical one-dimensional Mott-insulating charge transfer salt \cite{kaiser_optical_2014,singla_thz-frequency_2015}
and more recently reported as a microscopic mechanism for possible photomolecular superconductivity far above equilibrium $T_{\text{c}}$ in two-dimensional molecular crystals \cite{buzzi_photomolecular_2020,tindall_dynamical_2020}. The phononic route towards dynamical interactions is particularly suited for molecular crystals, where the effective local sites of the Hubbard model are comprised of entire molecules or parts thereof, which can be vibrationally excited and the on-site effective interaction thus changed [Fig.~\ref{fig:dynU}(b)] through the deformation of the intra-molecular wavefunction during the vibration. The key advantage of resonant phonon excitation in the mid-IR frequency range, compared to photodoping by optical pulses, lies in the spectral selectivity of the excitation pathway. This leads us to the discussion of vibrational control of emergent quantum phenomena in the following subsection.

\subsubsection{Nonlinear phononics}
\label{sec:phononics}
\begin{figure*}[t]
  \begin{center}
    \includegraphics[width=\textwidth,clip,page=9]{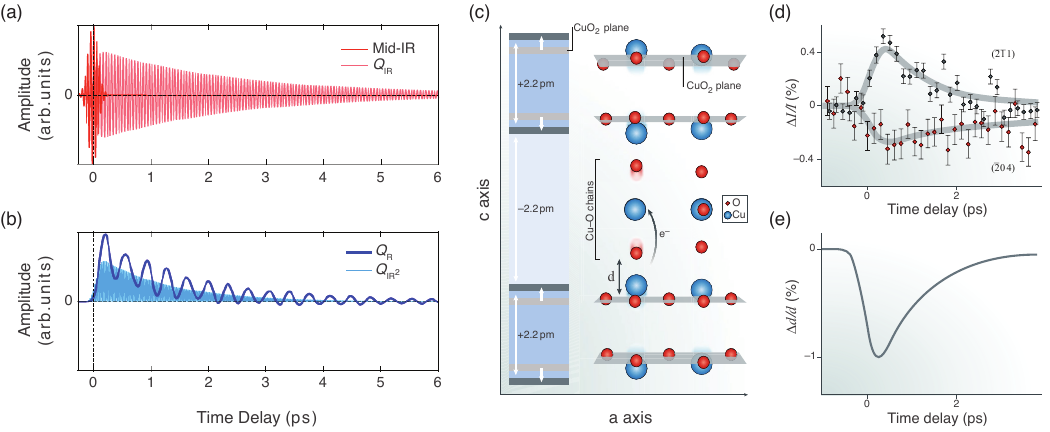}
  \caption{%{\bf 
  Transient crystal structure from nonlinear phononics. (a) Oscillatory dynamics of an IR-active phonon mode coordinate $Q_{\text{IR}}$ is induced under resonant driving by a mid-IR laser pulse. (b) Due to nonlinear phonon-phonon coupling to a Raman mode (coordinate $Q_{\text{R}}$) of the form $Q_{\text{IR}}^2 Q_{\text{R}}$, the potential energy of the Raman mode is transiently shifted to a new minimum, following the squared amplitude $Q_{\text{IR}}^2$. Accordingly, the Raman mode is displaced towards its transient potential minimum, which remains as long as the dynamics of the IR-active mode is coherent. (c) For YBa$_2$Cu$_3$O$_{6.5}$, the motion of the optically excited coherent $B_{1u}$ mode leads to a transiently modified crystal structure, which was measured using time-resolved X-ray diffraction \cite{mankowsky_nonlinear_2014} and is compatible with light-induced superconducting behavior. (d) Time-resolved relative change of intensity, $\Delta I/I$, of two representative Bragg peaks. Solid curves show the ab initio calculated structure based on nonlinear phonon-mode coupling. (e) Transient change in the Cu–O distance $d$ obtained from the calculation. Adapted from \onlinecite{mankowsky_coherent_2015} (a,b) and \onlinecite{buzzi_probing_2018} (c-e).}
  \label{fig:phononics}
  \end{center}
\end{figure*}

We now focus on resonant excitation of the crystal lattice with THz and mid-IR laser pulses. The notion of mode-selective control of the crystal lattice in quantum materials was coined `nonlinear phononics' and relies on the symmetries of the crystal and the phonon modes \cite{forst_nonlinear_2011,subedi_theory_2014}. There are three main differences from experiments where a pump interacts directly with the electronic degrees of freedom. (i) The energy from the pump is transferred into an initially coherent motion of an IR-active phonon mode, which then further interacts with other phonon modes, as well as, electrons and other collective degrees of freedom in the system. (ii) The targeted excitation of a particular phonon mode opens additional pathways for mode-selective control not accessible in traditional driving schemes. (iii) Salient effects can already happen \textit{during} the pump pulse and can often be understood as \textit{parametric excitation} of degrees of freedom that are coupled to the driven mode.

We consider a crystal with an inversion center in which parity is a good quantum number and phonon modes can be grouped into even (Raman-active) and odd (IR-active). Lowest-order anharmonicity leads to a nonlinear coupling of the form $g_{\text{cubic}} Q_{\text{IR}}^2 Q_{\text{R}}$ between an IR and a Raman (R) mode, with a coupling strength $g_{\text{cubic}}$ that is determined by the anharmonic potential. Coherent driving of the IR mode to large amplitudes [Fig.~\ref{fig:phononics}(a)], using strong mid-IR laser pulses, couples to the Raman modes such that a unidirectional displacive force occurs that is proportional to the square of the IR-mode displacement [Fig.~\ref{fig:phononics}(b)]. 
The consequences of the respective transient crystal structures [Fig.~\ref{fig:phononics}(c)] on the electronic subsystem have been shown to bear potential for changing the phases of quantum materials on ultrafast timescales \cite{mankowsky_non-equilibrium_2016}. As an example for fingerprints of a transiently modified lattice structure in pump-probe experiments, we show transient Bragg peak intensities [Fig.~\ref{fig:phononics}(d)] from femtosecond X-ray diffraction, supported by ab initio density functional theory calculations of the Cu-O distance [Fig.~\ref{fig:phononics}(e)], in the YBa$_2$Cu$_3$O$_{6.5}$ high-temperature superconductor \cite{mankowsky_nonlinear_2014}. 

Among others, coherent control of the crystal lattice by driving to anharmonic regimes has been suggested as a mechanism to induce ultrafast phase transitions, such as a lattice-controlled metal-insulator transition \cite{rini_control_2007}, and lead to states of matter without equilibrium counterparts, including possible light-induced superconductivity \cite{fausti_light-induced_2011,Hu2014,kaiser_optically_2014,mankowsky_nonlinear_2014}. In turn, these groundbreaking experiments have stimulated considerable theoretical activity on nonequilibrium superconductivity \cite{denny2015,raines_enhancement_2015,knap_dynamical_2016,kim_enhancing_2016,komnik_bcs_2016,okamoto_theory_2016,sentef_phonon-enhanced-SC_2016,sentef_theory_2017,kennes_transient_2017,sentef_light-enhanced_2017,coulthard_enhancement_2017,mazza_nonequilibrium_2017,murakami_nonequilibrium_2017,babadi_theory_2017,nava_cooling_2018,yao_wang_light-enhanced_2018,naga_eckstein_floquet_superconductivity_2018}, as well as a lively debate on how we should understand and interpret optical superconducting-like signatures on transient timescales \cite{demsar_nonequilibrium_2020,zhang_photoinduced_2020}. 
Furthermore, coherent optical driving of the lattice has been suggested as a path towards switching applications (see Sec.~\ref{sec:switching}) as well as a means of mapping of interatomic forces and potentials \cite{von_hoegen_probing_2018,Kozina2019}. Another developing subfield is phono-magnetism \cite{nova_effective_2017,afanasiev2019lightdriven,disa_polarizing_2020,juraschek_phono-magnetic_2020,stupakiewicz_ultrafast_2021,giorgianni_nonlinear_2021}, in which magnetic properties of materials are modified via coherent phonon driving.

To summarize, phononic control of quantum materials is a rapidly growing research field, conceptually located at the boundary between ultrafast dynamics after short laser excitation and dressed states of nonequilibrium matter, which will be discussed in the following.

\begin{figure*}[t]
  \begin{center}
    \includegraphics[width=\textwidth,clip,page=10]{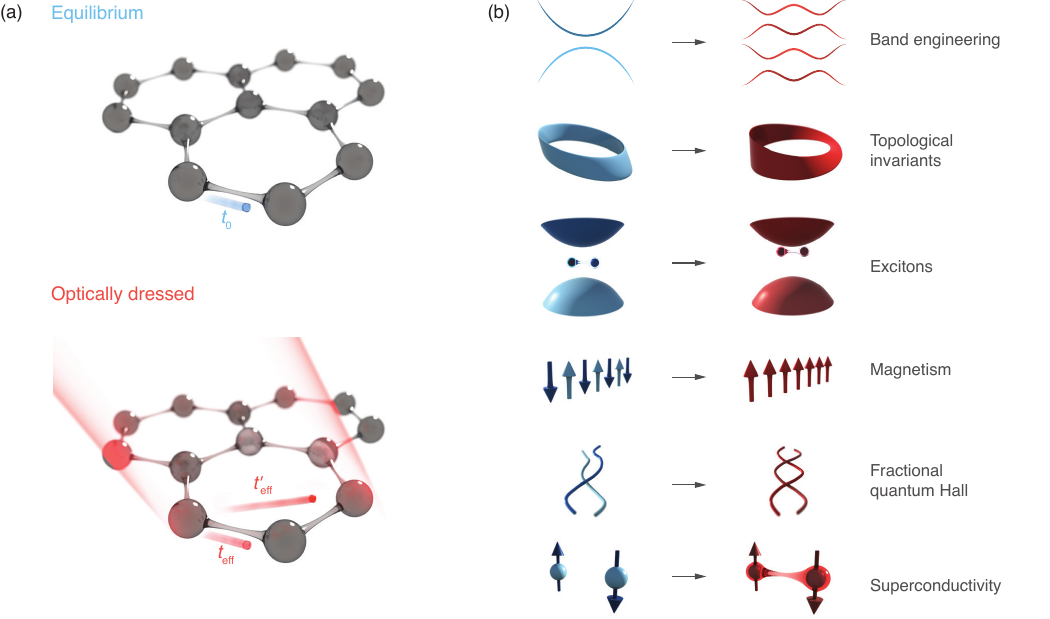}
  \caption{
  Versatile functionalities enabled by Floquet engineering. (a) Periodic driving using optical fields can steer the macroscopic properties of quantum materials by dressing their microscopic degrees of freedom, such as electronic hopping parameters. (b) Examples include proposals and experimental realizations of new electronic band structures \cite{GedikARPES1,GedikARPES2}, the manipulation of band topologies \cite{OkaAoki,Kitagawa,Lindner,McIver}, and the observation of excitonic Stark shifts caused by the creation of strong synthetic magnetic fields \cite{GedikStark1,WangStark}. Additional proposals include the optical control of magnetic correlations \cite{Mentink:Ultrafast,claassen17,Kennes:Floqueteng}, experimentally observed in quantum simulation settings \cite{Jotzu2}, the creation of fractionized quasiparticles obeying nonabelian statistics \cite{lee18}, and forms of photon-dressed superconductivity \cite{knap_dynamical_2016,Kennes:lightdwave,murakami_nonequilibrium_2017,tindall_dynamical_2020}.}  
  \label{fig:sec4}
  \end{center}
\end{figure*}

\section{Dressed states of nonequilibrium matter}
\label{sec:dressed}
Another class of nonequilibrium phenomena encompasses the effects that occur in solids \textit{during} optical illumination. Examples include high-harmonic generation \cite{HHG1,HHG2,HHGalex}, sub-cycle electron and spin dynamics \cite{ Subcycle1,Subcycle2,Subcycle3,Subcycle4}, light-field driven currents \cite{Lightfieldcurrent1,Lightfieldcurrent2} and the parametric amplification of plasma waves \cite{parametric1}. Comprehensive reviews of many of these topics can be found in \citealt{HHGreview} and \citealt{Nelsonreview}. Here, we focus on observables described by the coherent dressing of a material’s quantum states by the rapidly oscillating light-field. The manner and degree to which the states are dressed are determined by the amplitude, frequency and polarization of the electromagnetic drive. Such coherent light-matter interaction thereby provides an external control knob for transiently tuning the electronic and magnetic interactions in solids, which can be used to engineer quantum states with no equilibrium counterpart.

The consequences of photon dressing have long been investigated in the context of atomic systems and have led to applications in atomic clocks, precision spectroscopy, and quantum information processing \cite{Scully}. At the root of many of these phenomena is the alternating current (AC) Stark effect, a process whereby the oscillating electromagnetic field effectively shifts or splits electronic energy levels via the electric dipole interaction \cite{AutlerTownes}. This fundamental dressing effect has also been observed in solid state systems using the strong AC fields in ultrafast laser pulses \cite{Froelich,Rossi,Haug2008,Schfer2002,Bonitz2016,GedikStark1,reutzel_coherent_2020,Muecke}. 

Quantum materials can be dressed by light in more complex ways than isolated atomic systems by exploiting the additional degrees of freedom that interact with the light field. While this enhanced complexity is an opportunity for realizing novel material responses, it comes at the cost of increased dissipation, which leads to decoherence due to electron-electron, electron-phonon and other scattering mechanisms. Far stronger optical fields are thereby required to overcome these sources of decoherence in driven solid-state electronic systems \cite{GierzFloquet,Sato,Nuske}, which have historically made dressed states difficult to experimentally access.

Advances in the generation of strong-field laser pulses now make it possible to routinely access regimes where dressing effects dominate the response of optically driven quantum materials, which are being investigated with the aid of a new generation of ultrafast experimental probes \cite{GedikARPES1, McIver}. Theoretical breakthroughs utilizing Floquet theory \cite{Oka19Review,rudner_band_2020,Lindner} are simultaneously offering unprecedented guidance for experiments. This combined effort has led to remarkable discoveries and predictions (Fig.~\ref{fig:sec4}). We review some of these developments in the following sections.

\subsection{Theory introduction}
\label{sec:floquet}

\begin{figure*}[t]
  \begin{center}
   \includegraphics[width=\textwidth,clip,page=11]{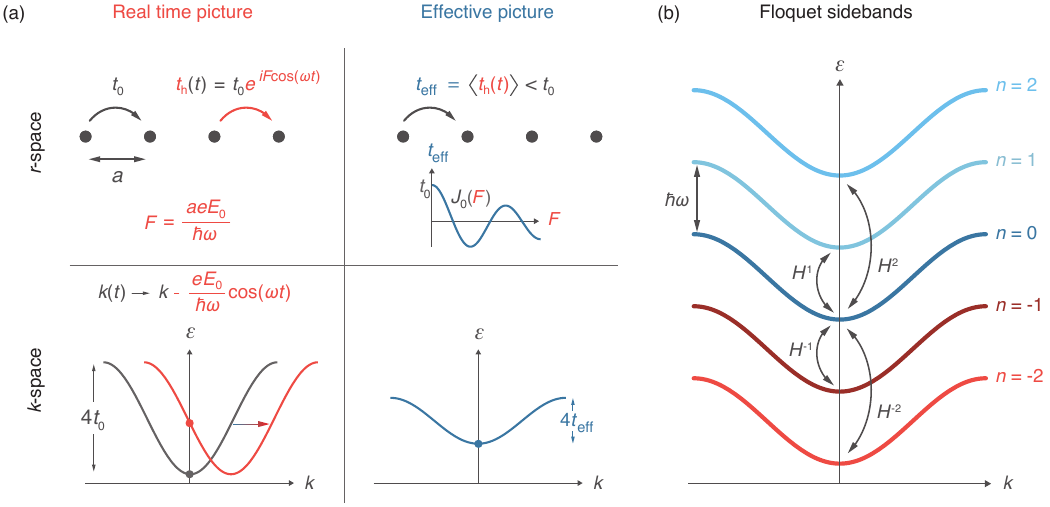}
  \caption{%\textbf{
  Dynamical localization in an atomic chain and Floquet sidebands. (a) Illustration of dynamical localization in an atomic chain visualized in real (upper row) and momentum space (lower row). The two columns represent the real time picture (left) as well as the effective, time-averaged picture (right) of the periodically driven system.
 When the equilibrium band dispersion (black, lower left subpanel) is driven back and forth in momentum space by the light field (red), the total bandwidth is reduced in the effective (dressed) picture (blue, lower right subpanel). This corresponds to a decrease in the effective hopping amplitude in real space, described by the Bessel function $t_{\rm eff}=t_0J_0(F)$ shown in the upper right sub-panel. (b) Visual representation of Floquet replicas according to Eq.~\eqref{eq:Fl_map}. At high frequency $\omega\to\infty$, the effective Hamiltonian is given by the time average~$H.^0$
  At finite frequency Floquet sidebands at energy distance $\omega$ become accessible via the higher moments $H^1$, $H^2$, $\dots$ of the Fourier series of the time dependent Hamiltonian $H(t)$ [the color coding of the sidebands corresponds to that of Eq.~\eqref{eq:Fl_map}].
  } 
  \label{fig:dynloc_big}
  \end{center}
\end{figure*}

To introduce the concept of dressed states in solids, we begin with the intuitive example of a periodically driven atomic chain in the high-frequency limit. We then generalize to the case of arbitrary frequency, for which Floquet theory provides a flexible and powerful framework to describe dressing effects.

\subsubsection{Dynamical localization in a driven atomic chain}

Consider a one-dimensional tight-binding model of noninteracting spinless fermions in second quantization
\begin{align}
    H=-\sum_n t_{0} c^\dagger_n c^{\phantom\dagger}_{n+1} +{\rm H.c.}%\label{eq:toy_1D_realspace}\\
     =\sum_k \epsilon(k) c^\dagger_k c^{\phantom\dagger}_{k}\,,%\label{eq:toy_1D_kspace}
\end{align}
where $t_{0}$ is the hopping amplitude and $c^{(\dagger)}_n$ annihilates (creates) a fermion on site $n$.  
In momentum space, one obtains a dispersion relation $\epsilon(k)=-2t_{0}\cos(ka)$, where $a$ is the lattice constant.

Next consider a coupling to a time-periodic longitudinal electric field $E(t)={E}_0\sin(\omega t)$ with a field strength ${E}_0$ and frequency $\omega$, which can be included by the Peierls substitution $t_{0}\to t_{ \rm h}(t) = t_{ 0} e^{i e a A(t) / \hbar}$ where $A(t)=E_0/\omega \cos(\omega t)$ is the vector potential, such that $E(t)=-\partial_t A(t)$. When the driving frequency is larger than the intrinsic energy and timescales in the system, the high-frequency oscillations cancel each other. Time averaging $t_{ \rm h}(t)$, denoted by $\left\langle \dots \right \rangle$, yields $t_{\rm eff}=\left\langle t_{\rm h}(t)\right\rangle=t_{0}J_0(F)$, where  $J_0$ is the $0^{th}$-order Bessel function. We introduced the dimensionless {\it Floquet parameter} 
\begin{equation}
    F=\frac{aeE_0}{\hbar\omega}, \label{eq:Floquetparameter}
\end{equation}to parameterize the strength of the electromagnetic drive, with $e$ the elementary charge. Since $\left|J_0(x)\right|<1$, the resulting effective hopping amplitude $t_{\rm eff}$ is always smaller than the equilibrium value $t_{\rm 0}$, meaning that on average (i.e., in the effective or dressed picture) the electrons in the driven chain are more localized than in the undriven chain. This real-space description of \emph{dynamical localization} is depicted in Fig.~\ref{fig:dynloc_big}(a).

Equivalently, in momentum space, the effect of dynamical localization is captured by the minimal coupling  $k\to k-eA(t)/\hbar$ in the band dispersion. The effect of the electric field is to periodically \emph{shake} the dispersion to the left and right [bottom row of Fig.~\ref{fig:dynloc_big}(a)]. When considering the minimum of the equilibrium dispersion relation at $k=0$, one finds that this shaking on average shifts the bottom of the band dispersion to higher energy in the effective picture. Analogously, the dispersion maxima at the zone boundaries are lowered in energy upon averaging. The net effect in the effective, or \emph{dressed}, picture is a reduction in bandwidth compared to the equilibrium dispersion relation. This example demonstrates how periodic driving provides a tuning knob for materials properties via the renormalization of electronic hopping  \cite{Dunlap:DynLoc,Bucksbaum:H2softening,claassen17,Mentink:Ultrafast,Kennes:lightdwave,Kennes:Floqueteng}.

\subsubsection{Floquet primer} 
Floquet theory provides a powerful basis to treat a periodically driven quantum system beyond high- or low-frequency expansions. Such a system is governed by the time-dependent Schr\"odinger equation:
\begin{equation}
    i\hbar \frac{d}{dt}\left|\Psi(t)\right\rangle=H(t)\left|\Psi(t)\right\rangle,
\label{eq:Floquet}
\end{equation}
with a time-periodic Hamiltonian $H(t+T)=H(t)$ with period $T$. 
\citealt{Floquet83} showed that the solution to Eq.~\eqref{eq:Floquet} can be written as:
 \begin{equation}
 \left|\Psi_l(t)\right\rangle= e^{-i\epsilon_lt/\hbar}\left|\Phi_l(t)\right\rangle,
 \end{equation}
 where the $T$-periodic wavefunction $\left|\Phi_l(t)\right\rangle=\left|\Phi_l(t+T)\right\rangle$ and the quasi-energies $\epsilon_l$ are determined by the algebraic equation 
 \begin{equation}
     \left(\epsilon_l+n\hbar\omega\right)|\Phi_l^n\rangle= \sum_{n'}H^{n-n'}|\Phi_l^{n'}\rangle,
 \end{equation}
 with frequency $\omega=2\pi/T$ and Fourier decompositions $H^n=\frac{1}{T}\int_0^T dt e^{in\omega t}H(t)$ and $\left|\Phi_l^{n}\right\rangle=\frac{1}{T}\int_0^T dt e^{in\omega t}\left|\Phi_l(t)\right\rangle$. This maps a periodic differential equation to an algebraic (quasi-equilibrium) problem. The time-dependent problem is transformed into an effective static Hamiltonian by considering $\underline\Phi_l=\begin{pmatrix}\dots&
 \left|\Phi_l^{-1}\right\rangle&
 \left|\Phi_l^{0}\right\rangle&
 \left|\Phi_l^{1}\right\rangle&
 \dots
  \end{pmatrix}^T
 $ and  $ \mathcal{H} \underline\Phi_l=\epsilon_l\underline\Phi_l$, such that

  \begin{equation}
 {\footnotesize\mathcal{H}=\begin{pmatrix}
 \ddots &H^{-1}&H^{-2}&H^{-3}&H^{-4}&&&\\
 H^{1}&{\color{colorp1}H^{0}-(-1)\hbar\omega}&H^{-1}&H^{-2}&H^{-3}\\
 H^{2}&H^{1}&{\color{color0}H^{0}-0\hbar\omega}&H^{-1}&H^{-2}\\ H^{3}&H^{2}&H^{1}&{\color{colorm1}H^{0}-1\hbar\omega}&H^{-1}\\
 H^{4}&H^{3}&H^{2}&H^1&\ddots\\
 \end{pmatrix}}.
  \label{eq:Fl_map}
 %\;\;\;\;\; 
 \end{equation}

\begin{figure*}[t]
  \begin{center}
  \includegraphics[width=\textwidth,clip,page=12]{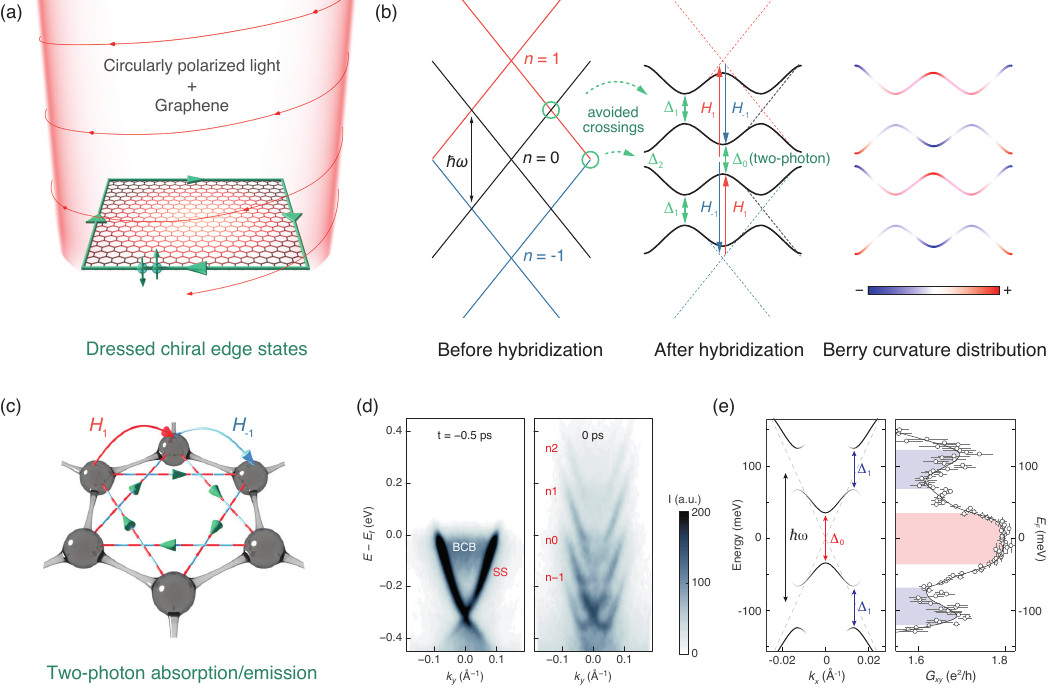}
%  \vspace{2pt}
  \caption{%\textbf{
  Topological Floquet engineering in Dirac systems. (a) A coherent interaction with circularly polarized light was predicted to induce a photon-dressed topological band structure in graphene, characterized by protected chiral edge states \cite{OkaAoki,Kitagawa}. (b) When the Floquet sidebands form they hybridize at their crossing points, opening band gaps ($\Delta_{1}$, $\Delta_{2}$). A gap also opens at the Dirac point ($\Delta_{0}$) due to a two-photon absorption/emission process that breaks time-reversal symmetry. The resulting Floquet-Bloch bands are fully gapped and possess a nonzero total Berry curvature and Chern number. (c) Real space schematic of the two-photon absorption/emission hopping process that induces a gap at the Dirac point in graphene. (d) tr-ARPES data of Bi$_{2}$Se$_{3}$ detecting the formation of Floquet-Bloch bands from the Dirac surface states. Taken from \citealt{GedikARPES2}. (e) Nonequilibrium anomalous Hall conductance of graphene driven by circularly polarized light as a function of the equilibrium Fermi level ($E_{F}$). Red and blue shading corresponds to $E_{F}$ being gated to the regions where the gaps $\Delta_{0}$ and $\Delta_{1}$ were predicted to appear. Taken from \citealt{McIver}.} 
  \label{fig:topo}
  \end{center}
\end{figure*}

$\mathcal{H}$ consists of an infinite number of copies of the time-averaged Hamiltonian $H^0$ energetically shifted by integer multiples of the driving frequency  $\omega$. The higher harmonics $H^1$, $H^2$, $\dots$ couple the nearest, next-nearest, $\dots$ neighboring copies in this effective language. In the context of solids, where the undriven system is described by Bloch bands, these copies, or Floquet replicas, lead to Floquet-Bloch bands repeating periodically in momentum (Bloch) as well as energy (Floquet) space [Fig.~\ref{fig:dynloc_big}(b)].  
This insight results in the notion of sidebands emerging under periodic driving, which can be used to alter the physical properties of a system dramatically \cite{OkaAoki,rudner_band_2020,Kennes:solid} (Fig.~\ref{fig:topo}). 
Although Floquet theory was introduced more than 100 years ago, its implementation within new theoretical approaches (Sec.~\ref{sec:theory_advances}) is a subject of intense research. Green's functions, other diagrammatics, or effective single-particle pictures, such as used in dynamical mean-field theory, functional renormalization group, or density functional theory, can rather easily exploit these ideas \cite{giovannini_simulating_2013,Eissing2016b,aoki_nonequilibrium_2014,PhysRevB.78.235124}. For tensor networks \cite{Eggert:FDMRG,Kennes:Floqueteng} or other variational techniques this is substantially complicated by the fact that Floquet theory requires the determination of the full spectrum. 

Floquet engineering is challenging in quantum materials owing to the large optical fields ($10^{7}-10^{8}$ V/m) and long wavelengths needed to obtain a sizeable Floquet parameter (Eq.~\eqref{eq:Floquetparameter}). This necessitates the use of intense ultrafast laser pulses and limits current experiments to the low-frequency regime. While Floquet theory inherently assumes continuous wave driving, experimental results obtained using laser pulses with multiple optical cycles have nevertheless been well-described using this framework \cite{GedikARPES1,GedikARPES2,McIver,GedikStark1,WangStark}. We review some of these experimental results in the following. 

\subsection{Dressed band structures: Floquet topological insulators} 
\label{sec:bandstructure}

Optically dressed quantum materials garnered considerable interest after they were proposed capable of hosting topologically nontrivial states that could be manipulated by adjusting the light field \cite{OkaAoki,Kitagawa,Lindner}. Such systems are often referred to as `Floquet topological insulators', and in some cases they are characterized by topological invariants that have no equilibrium counterpart \cite{Kitagawa2,Rudner}. 

The first proposal along these lines was put forward by \citealt{OkaAoki}. Using Floquet theory it was shown that an anomalous Hall effect could be induced in graphene by dressing its quantum states with circularly polarized light [Fig.~\ref{fig:topo}(a)] \cite{Oka19Review}. The physics can be summarized as follows: Floquet-Bloch bands form and hybridize at their crossing points, opening band gaps [$\Delta_1$, $\Delta_2$ in Fig.~\ref{fig:topo}(b)]. A second-order process also causes a gap to open at the Dirac point ($\Delta_0$) because the interaction breaks time-reversal symmetry. The resulting fully-gapped Floquet-Bloch bands have a nonzero Berry curvature distribution that causes charges to undergo an anomalous Hall effect. By reversing the helicity of the light, the polarity of the induced Berry curvature distribution, and hence the direction of the net anomalous Hall current, can be controlled. 

In the high-frequency limit, where the Floquet sidebands do not overlap and thus there are no gaps due to band crossings, the Hamiltonian for the driven system is
\begin{equation}
     H_{\text{eff}} \approx H_{0} + \frac{[H_{-1},H_{1}]}{\hbar\omega}, \label{eq:Aoki}
 \end{equation}
where the first term is the equilibrium Dirac Hamiltonian and the second term is responsible for the gap opening at the Dirac point. Specifically for a two-dimensional Dirac cone with energy-momentum relation $\epsilon(\vec{k}) = \hbar v_F |\vec{k}|$, with Fermi velocity $v_F$, one obtains $\Delta_0 = 2\hbar^{-1} e^2 v_F^2 E_0^2/\omega^3 = 2\hbar^{-1} e^2 v_F^2 A_0^2/\omega$, which scales quadratically with the peak electric field $E_0$ and is enhanced for small frequencies.

\citealt{Kitagawa} interpreted Eq.~\eqref{eq:Aoki} as the energy difference ($\Delta_0$) between two processes, where an electron in the $n=0$ Dirac cone first absorbs (emits) a circularly polarized photon to enter the $n=1$ ($n=-1$) sector, then emits (absorbs) a photon of the same helicity to return to the $n=0$ sector [Fig.~\ref{fig:topo}(b)]. In real space, this corresponds to a double hopping in the unit cell that has the net effect of introducing chiral intra-sublattice tunneling elements to the effective Hamiltonian [Fig.~\ref{fig:topo}(c)]. The effective Hamiltonian is identical to that proposed by Haldane for a topological Chern insulator in graphene \cite{Haldane}, which was previously believed to be inaccessible via any realistic experimental setting. When only the lower Chern band is populated, a topological phase is expected to develop hosting a quantized anomalous Hall effect carried by chiral edge states, which would be photon-dressed in this case. 

Floquet topological insulators have been studied in the high-frequency limit using a variety of quantum simulation platforms, most notably in photonic waveguides \cite{Rechtsman}, where the propagating edge mode was directly observed, and in driven optical lattices of ultracold fermions \cite{Jotzu}. Later theoretical works confirmed that topological states could also be created in the low-frequency limit [Fig.~\ref{fig:topo}(b)], with the difference being that the multitude of Floquet-Bloch bands that form due to the additional gap openings can have Chern numbers $|C| > 1$, i.e. , multiple edge modes \cite{Usaj,Mikami,Mitra,Sentef,Claassen:All}. An additional topological invariant, dubbed winding number, can also arise in the low-frequency limit \cite{Kitagawa2,Rudner}, which is defined over time rather than momentum space and, thus, only realizable in periodically driven systems. Observations of such `anomalous Floquet topological insulators' have also been reported in the quantum simulation community \cite{Kitagawa3,Maczewsky}.   

In a landmark paper, \citealt{GedikARPES1} reported the first spectroscopic evidence of topological Floquet-Bloch bands in a solid within the Dirac surface states of the topological insulator Bi$_2$Se$_3$ using tr-ARPES [Fig.~\ref{fig:topo}(d)]. The data reveal multiple dressed sidebands and the opening of band gaps, similar to those illustrated in [Fig.~\ref{fig:topo}(b)].

\citealt{McIver} reported the first electrical transport results from topological Floquet-Bloch bands. Using an ultrafast transport device  (bottom center panel of Fig.~\ref{fig:temp1}), an anomalous Hall effect was detected in graphene illuminated by a circularly polarized mid-IR pulse. Remarkably, at high laser pulse fluences they observed that the anomalous Hall conductance saturated on the order of two conductance quanta\textemdash the same value predicted in the high-frequency limit \cite{Kitagawa}. When the graphene Fermi level was tuned away from charge neutrality using an electrostatic gate [Fig.~\ref{fig:topo}(e)], features were observed in the conductance spectrum closely aligned with the opening of the band gaps $\Delta_0$ and $\Delta_1$, as predicted by Floquet theory for their laser pulse parameters. Subsequent numerical studies confirmed that the observed photocurrents originated from the formation of topological Floquet-Bloch bands \cite{Sato,Nuske}. These results are an encouraging sign that Floquet-engineered topological edge states are a distinct possibility in optically driven Dirac materials.

\begin{figure}
    \centering
    \includegraphics[width=\columnwidth, clip,page=13]{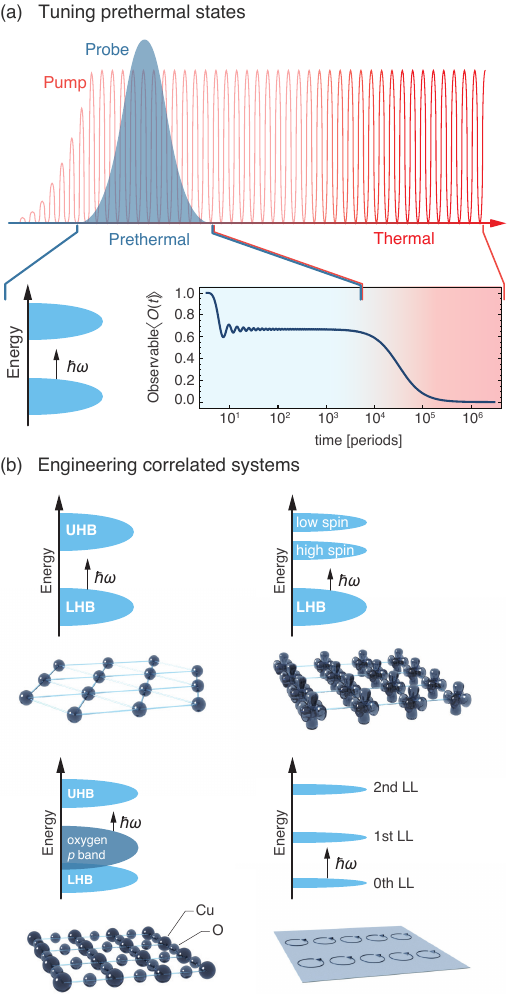}
    \caption{%{\bf 
    Routes to avoid runaway heating. (a) Generic interacting systems heat up upon applying a continuous drive, detrimental to coherent control. However, by tuning the drive frequency to a spectral gap in the system, this heating can in principle be pushed to (exponentially) large times. Control of prethermal states can then be achieved on intermediate to long timescales until thermalization sets in. (b) Examples of condensed matter systems for which such a strategy could be applied. The upper two and bottom left subpanels show Mott and charge transfer insulators (LHB and UHB denote the lower and upper Hubbard band, respectively). Another example are quantum Hall insulators (lower right subpanel). Between the different Landau-Levels (LL) energy gaps proportional to the externally applied magnetic field emerge.  Frequency detuning to the sharp resonances between LLs could, in principle, allow for prethermal state Floquet control of the system.
  }
    \label{fig:corrFloquet}
\end{figure}

\subsection{Towards Floquet many-body physics: Heating and interactions}
\label{sec:heating}

The experiments by \citealt{GedikARPES1} and \citealt{McIver} demonstrate that Floquet engineering in quantum materials is realistic despite the presence of strong dissipation, which leads to decoherence. However, the \mbox{tr-ARPES} data also reveals significant pump-induced carrier excitation, leading to a nonequilibrium dressed electron distribution in the Floquet-Bloch bands. Here, distributional and dissipative effects play an important role already even in the absence of electronic interactions \cite{dehghani14,seetharam15,iadecola15}, and contribute for instance to substantial deviations from quantization of the Hall conductivity and edge state transport in Floquet band topological insulators \cite{kundu13,farrell16}. The roles of heating and thermalization hence can be expected to take on an outsized role as `Floquet engineering' is extended towards the control of interacting many-body states of matter. Therefore, materials and driving protocols need to be chosen carefully in order to minimize heating and carrier redistribution such that a nonequilibrium state proximal to the ground state of the effective Hamiltonian is realized. 

Theoretically, there have been multiple proposals to address the challenge of heating. In general, closed periodically driven interacting systems continuously absorb energy and heat to a featureless infinite-temperature steady state by virtue of the eigenstate thermalization hypothesis for nonintegrable systems \cite{dalessio14}.
However, tailored drive protocols can be chosen to suppress heating at short times and realize a potentially long-lived `prethermal' regime. Here, a central ingredient is a separation of energy scales, and, thus, a separation of timescales in the driven system, which entail an ergodic obstruction to fast energy absorption. After an initial ramp-on period, governed by fast switch-on processes, the intermediate time (period-averaged) dynamics can saturate to a prethermal plateau, governed by an effective Hamiltonian that captures a controlled Floquet modification, leading to observable consequences for instance in correlation functions or spectroscopic probes of the system. This regime is bounded by a timescale that signifies the onset of heating towards a featureless high-temperature state [Fig.~\ref{fig:corrFloquet}], which can in principle nevertheless retain certain correlations when constrained by exact nonabelian symmetries \cite{tindall19}.

Most of our current understanding regards the limit of high-frequency driving \cite{machado19a,machado19b,gulden19,haldar18,szabolcs18,Peronaci:Resonant,weidinger17,claassen17,Mentink:Ultrafast}, where the pump frequency exceeds local energy scales. In this limit rigorous results for slow heating were recently established \cite{abanin15,mori16,kuwahara16,ho18,abanin17a,abanin17b,mori18}. Notably, these works provide only mathematical bounds for the onset of heating to featureless states at infinite temperature, which can be exceeded in specific settings. 
%and a high-frequency regime is generally absent in real materials, it is nevertheless expected that certain solid state systems realize such regimes in specific settings.
Many-body localized (MBL) systems constitute an important exception \cite{lazarides14,dalessio13,ponte15,lazarides15}. Here, heating can be averted via a lack of ergodicity, stabilizing a tantalizing array of \textit{bona fide} nonequilibrium phases, such as time-crystalline orders or Floquet topological phases \cite{rovny18,zeng17,Decker:FloquetTopo}. 

In practice, a high-frequency driving limit is absent in most real materials, since higher-lying bands and collective  excitations provide a multitude of possibilities for resonant absorption at higher frequencies. Similarly, while MBL is realizable in cold atomic systems \cite{singh19}, it is typically destabilized in solids due to energy dissipation to the lattice or other degrees of freedom. Nonetheless, sufficiently slow energy absorption, for instance due to off-resonant driving or other ergodic obstructions to heating, can realize `prethermal' dynamical regimes at short times (Fig.~\ref{fig:corrFloquet}), which can harbor intriguing emergent phenomena \cite{haldar18,gulden19,luitz19,rovny18,claassen21}. Therefore, a central problem is to identify materials and driving regimes for which comparatively long-lived prethermal regimes persist in a realistic setting. %For instance, while it is currently impossible using the current generation of Dirac materials, however, advances in van der Waals heterostructure design may make it possible to create  such a system \cite{vdWreview} in the future. 
%Thus, currently, a main goal of Floquet engineering in solids is to exploit prethermal dynamics to engineer the Hamiltonian dynamics and steer a correlated system on short timescales.
Notably, these arguments imply that Floquet control of \textit{quantum phases} in solids entails a two-fold challenge: engineering (i) a prethermal Hamiltonian that captures the desired dynamics on sufficiently long timescales, and (ii) an electronic distribution with respect to this prethermal Hamiltonian on the same timescales. 

A complementary route to Floquet engineering 
aims to exploit dissipation in {\it open} systems to stabilize a driven steady state at long times, for which the energy influx absorbed from the pump is compensated by energy dissipation into the environment \cite{Seetharam19Floquet,Esin18}. In a solid-state setting, the crystal lattice can act as a `thermostat' for the electronic system due to the large timescale separation between electronic and lattice dynamics, with early works suggesting routes towards the controlled dissipative population of single-particle Floquet states \cite{seetharam15,dehghani14,iadecola15}. The phase diagrams of infinite-time steady states of clean systems have been established to exhibit rich phase transitions \cite{Peronaci:Resonant,Jose:Ultra,Mitra:Nonequilibrium,Walldorf:AFM, kalthoff_nonequilibrium_2021,PhysRevLett.125.147601}, which are, however, expected to be first order at finite driving frequency \cite{PhysRevLett.122.110602}.

\subsection{Engineering correlated systems}
\label{sec:engineering}

A tantalizing prospect concerns utilizing tailored light pulses to engineer novel phases of matter in correlated electron systems. %Here, the nonperturbative interplay of prethermalization and dissipation at short and intermediate timescales promises a rich playground to stabilize new phases of matter but remains a largely unexplored and methodologically challenging regime. We discuss in the following different examples.
The use of light to modulate interactions or selectively break symmetries to tune the interplay of competing phases promises a rich playground to stabilize new phases of matter but remains a largely unexplored and methodologically challenging regime. We discuss in the following different examples.
%a rich playground to both stabilize unconventional nonequilibrium states of matter and to probe the role of competing interactions. 
%However, in contrast to Floquet engineering of single-particle band structures discussed above which does not rely on controlling the carrier distribution, the necessity of strong fields and resonant drives will lead to rapid heating in interacting systems, masking any signatures of controlled manipulation in realistic materials. An exception concerns systems with exact nonabelian symmetries, in which certain correlations persist even at infinite temperatures \cite{tindall19}.

\begin{figure*}
    \centering
    \includegraphics[width=\textwidth,page=14]{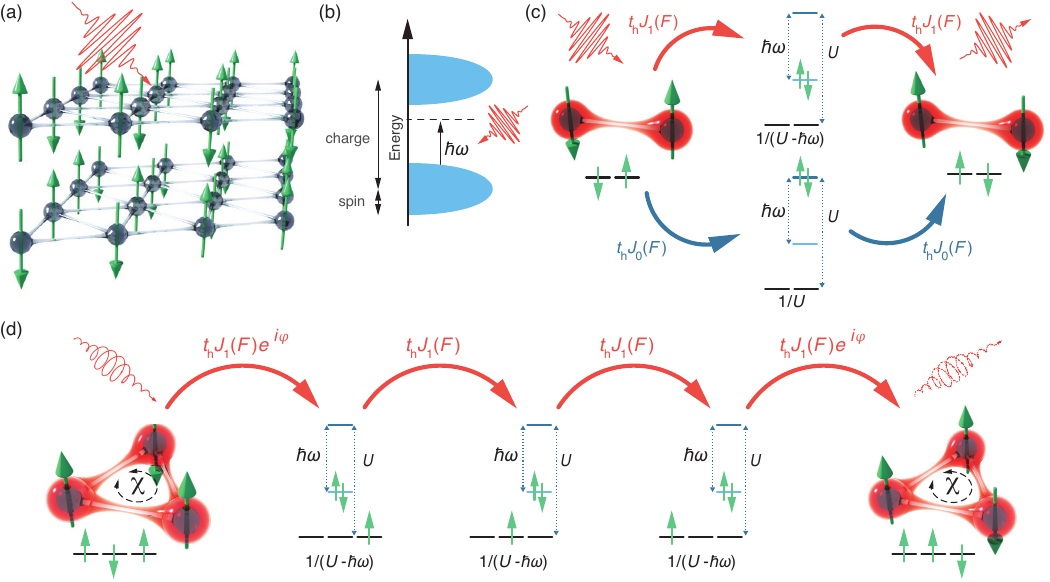}
    \caption{%{\bf 
    Floquet engineering of magnetic interactions. (a) Schematic of manipulating magnetic degrees of freedom via irradiation of quantum magnets with light. (b) In Mott insulators, pumping below the charge gap suppresses charge excitations on short timescales while renormalizing the magnetic interactions at low energies. (c) Irradiation with light can alter conventional spin-exchange interactions via reducing the energy cost of intermediate virtual states of electronic tunnelling. (d) Circularly polarized light can induce new types of magnetic interactions, such as SU(2)-symmetric chiral three-spin exchange interactions due to dynamical time-reversal symmetry breaking, which are absent in equilibrium.}
    \label{fig:magnetism}
\end{figure*}

{\it Magnetic Mott insulators} have recently emerged as a promising class of candidate materials for Floquet engineering in correlated systems  \cite{Mentink:Ultrafast,bukov15b,claassen17,Walldorf:AFM,Kennes:Floqueteng}. In a Mott insulator, strong local Coulomb repulsion---typically in localized transition-metal $d$ orbitals---freezes local charge degrees of freedom at commensurate filling to open an insulating gap and form local magnetic moments. If the pump frequency is sufficiently red-detuned from the charge gap, energy cannot be absorbed resonantly as photons couple to charge [Fig.~\ref{fig:magnetism}(a,b)]. Remarkably, while the charge sector remains inert on short timescales, a prethermal regime can emerge which is characterized by transiently modified effective magnetic interactions between the local moments \cite{Mentink:Ultrafast,claassen17}.

A minimal model to describe these effects is the driven half-filled single-band Hubbard model
\begin{align}
    \hat{H}(t) = -t_{\rm h} \sum_{\left<ij\right>\sigma} e^{i e \mathbf{A}(t) \cdot \mathbf{r}_{ij} / \hbar}~ \hat{c}^\dag_{i\sigma} \hat{c}^{\vphantom{\dag}}_{j\sigma} + U \sum_{i} \hat{n}_{i\uparrow} \hat{n}_{i\downarrow},
\end{align}
where $t_{\rm h}$ and $U$ denote hopping of electrons between neighboring sites and local Coulomb repulsion, respectively. The optical pump enters via minimal coupling to a gauge field $\mathbf{A}(t)$. 

In equilibrium, the half-filled Hubbard model at low energies and large $U$ canonically maps onto effective nearest-neighbor Heisenberg interactions $J \approx \frac{4 t_{\rm h}^2}{U}$ which arise from virtual exchange. Suppose that the pump frequency $\omega$ is sufficiently red-detuned from the Mott gap $\sim U - 4t_{\rm h}$. If the charge sector remains inert out of equilibrium, then a perturbative calculation, which simultaneously integrates out charge degrees of freedom and photons, leads to a transient photoinduced \textit{renormalization} of spin-exchange interactions, proportional to $J \approx \sum_m \frac{4 t_{\rm h}^2 [J_m(F)]^2}{U - m\hbar\omega}$ \cite{Mentink:Ultrafast,PhysRevLett.115.075301,PhysRevB.100.220403}. 

A physical picture readily emerges by noting that the electron participating in the exchange process can now lower the Coulomb repulsion energy $U$ of the virtual intermediate state by absorbing a photon from the pump field, thereby enhancing the effective spin exchange. A schematic depiction of this process is shown in Fig.~\ref{fig:magnetism}(c). Conversely, for strong pump fields, higher-order multi-photon processes can dominate and even permit flipping the sign of Heisenberg exchange if $m$-photon processes with $m\hbar\omega > U$ dominate \cite{Mentink:Ultrafast}. For a typical Mott antiferromagnet with $U \sim 2$~eV, \mbox{$t_{\rm h} \sim 0.1$~eV}, $a \sim 3\textrm{~\AA}$ (corresponding to $J \sim 20$~meV in equilibrium), and pumping at $700$~nm, one finds an enhancement of the exchange coupling in the range $\Delta J \sim 0.2$ to 20
~meV for a peak field strength of 0.05 to $0.5 \textrm{~V \AA}{}^{-1}$ \cite{Batignani2015}.

{\it Frustrated quantum magnets}, encompassing materials in which competing magnetic interactions render minimizing the free energy landscape nontrivial, are another interesting class to study. In these compounds, transient modifications of the magnetic interactions are expected to have a sizeable effect. While the role of overall renormalization of Heisenberg interactions discussed above is reflected solely in shifts of the excitation spectrum without affecting the ground state, strongly frustrated systems typically host rich phase diagrams of competing magnetic orders such that a small photoinduced transient change of competing \textit{subdominant} interactions can `nudge' the system into a proximal phase \cite{Kennes:lightdwave}. Moreover, such materials are highly sensitive to the breaking of symmetries, which grants a powerful handle to steer the nonequilibrium phase on prethermal timescales, which can be kept sufficiently long via detuning from the Mott gap.

From these considerations, one avenue of utilizing Floquet engineering to transiently stabilize a novel correlated state of matter builds upon the observation that pumping a frustrated Mott insulator with circularly polarized light
permits the selective breaking of time-reversal symmetry (TRS) and inversion without breaking the SU(2) spin-rotational symmetry of the magnetic moments \cite{claassen17,PhysRevB.96.014406}. Notably, while TRS can be readily broken in equilibrium via external magnetic fields, the dominant contribution to low-energy spin dynamics comes from a Zeeman splitting that breaks SU(2), quickly magnetizing the system. Instead, the role of the pump is to generate a transient photoinduced scalar spin chirality  $\mathbf{S}_i \cdot (\mathbf{S}_j \times \mathbf{S}_k)$, whose form can be inferred from symmetry considerations alone, and microscopically arises from fourth-order virtual processes involving elementary triangles of the underlying lattice  [Fig.~\ref{fig:magnetism}(d)]. For example, it was shown that in Kagome antiferromagnets, such as herbertsmithite, a weak optical pump tuned between one- and two-photon resonances of the parent Mott insulator can induce a transition to a proximal chiral spin liquid phase \cite{claassen17}, suggesting a novel nonequilibrium pathway to one of the earliest and most elusive candidates of a gapped quantum spin liquid with topological order.

{\it Strong spin-orbit coupled multi-orbital compounds} constitute a natural generalization of such a nonequilibrium separation of charge and spin degrees of freedom \cite{liu18,hejazi18,arakawa21,sriram21}, which could allow for Floquet engineering while mitigating heating. In the orthorhombic titanates, the partially-filled $t_{2g}$ manifold has been proposed as a prime target for optical manipulation of effective Kugel-Khomskii spin-orbital interactions. In Kitaev materials such as $\alpha$-RuCl$_3$, Floquet engineering can modify the interplay between Kitaev and competing exchange interactions to manipulate the magnetic state and induce a quantum spin liquid. In charge transfer insulators such as the cuprates, photoinduced rotational symmetry breaking was proposed to engineer magnetic interactions and ultimately destabilize the Mott antiferromagnetic phase in favor of a $d$-wave superconductor \cite{Kennes:lightdwave}. Conversely, theoretical predictions of spectroscopic signatures suggest that the controlled manipulation of magnetic exchange interactions could be observed in time-resolved Raman or resonant inelastic X-ray scattering \cite{wang18}, for instance via tracking the softening of bimagnon excitations \cite{PhysRevB.98.245106}.

{\it Fractional quantum Hall systems} are another intriguing direction to engineer prethermal correlated phases using external driving.
While a vast series of Abelian filling fractions have been experimentally realized \cite{RevModPhys.71.S298}, more elusive fractional quantum Hall states \cite{mooreread,PhysRevB.59.8084} with non-Abelian quasiparticles are well-known to prefer fine-tuned or multi-body interactions for stabilization \cite{PhysRevLett.66.3205}, while simultaneously incurring tremendous interest for potential applications in topological quantum computing and beyond \cite{RevModPhys.80.1083}. In principle, Landau levels in ultraclean electron gases provide a natural separation of energy scales for Floquet engineering. If the pump frequency is sufficiently detuned from a cyclotron resonance, the regularity in Landau level spacing guarantees the absence of multi-photon resonances at low photon numbers. If heating can be suppressed, such a setting has been predicted to give rise to photoinduced three-body interactions that arise from virtual inter-Landau-level scattering \cite{lee18}, suggesting a nonequilibrium route to stabilize a Moore-Read state. Conversely, resonant excitation between Landau levels has been proposed to engineer effective multi-layer fractional quantum Hall models \cite{ghazaryan16}.

\section{Outlook}
\label{sec:outlook}

In this Colloquium, we discussed a set of results that provide insights into ultrafast phenomena in photoexcited quantum materials. These advances rely on an unprecedented range of optical excitations and a growing set of tools to probe and simulate nonthermal pathways towards control and functionality on ultrafast timescales. In particular, photoinduced phase transitions to thermally inaccessible states via short optical excitation or photon dressing constitute promising routes towards achieving nonequilibrium functionality. 

Recent studies of photoinduced phase transitions have provided a more detailed understanding of the relevant microscopic degrees of freedom and their nonthermal behavior, and have revealed the interplay between these degrees of freedom in the time domain. The realization of new applications based on optical switching into metastable phases is already being explored. For instance, the photoinduced charge configuration change of 1\textit{T}-TaS$_2$ emerges as a new avenue for memory devices in low-temperature circuitry \cite{MihailovicCCM}. Similarly, advances in optical switching of superconductors \cite{budden_evidence_2021} could be integrated in future microelectronics. In order to expand the range of applications, mandates identifying further materials whose nonthermal energy landscape allows for efficient switching mechanisms into metastable states.   

Creation and control of properties through light-matter coupling via Floquet engineering is another promising approach. In order to realize robust functionality and devices, it is necessary to understand the role of dissipation and identify strategies to mitigate heating and decoherence. As such, a firm understanding of the coupling to environmental degrees of freedom is necessary. Moreover, suitable candidate materials need to be identified which host appropriate electronic structures and interactions to minimize energy absorption at selected frequencies while still permitting a controlled modification of the salient electronic features. 

In order to reach the final goal of \textit{new functionalities} based on quantum dynamics far from equilibrium, it will be of increasing importance to (i) design materials tailored towards enabling precise nonequilibrium control by integrating materials synthesis and nonequilibrium experimentation synergistically, and to (ii) establish further bridges between the existing ultrafast material science community and adjacent research fields. For example, the young field of polaritonic chemistry is bridging nonequilibrium quantum chemistry, quantum optics, and nanoplasmonics \cite{ebbesen_hybrid_2016,feist_polaritonic_2018,ruggenthaler_quantum-electrodynamical_2018,flick_strong_2018}. Inspired in part by these efforts, \emph{cavity material science} is an emergent topic at the boundary between quantum optics and nonequilibrium material science \cite{juraschek_cavity_2019,hubener_engineering_2020}. The key idea is the replacement of a strong laser by a few-photon state in a strongly light-matter-coupled cavity that enables control of the material properties. One possible route towards cavity control could employ the quantum nature of photons for effective Floquet engineering without the need for intense laser fields \cite{sentef_crossover_2020}.

 %Similarly, control of electron-phonon couplings and superconductivity in nanophotonic architectures was theoretically investigated \cite{sentef_cavity_2018} and recently explored experimentally \cite{thomas_exploring_2019}.

Controlled synthesis of van der Waals heterostructures has recently emerged as a versatile tool to establish electronic and structural properties with unprecedented control and variety. In this context, twistronics---Moir\'e potential engineering by twisting adjacent layers---is set to play a crucial role. The twist angle provides a critical handle on designing electronic and structural properties which can be geared with great flexibility \cite{moireqs}. Crucially, the relevant kinetic energy scales of the Moir\'e superlattice can be orders of magnitude smaller than in the bulk material, suggesting that external nonequilibrium perturbations can have an outsized effect in determining electronic phases. Combining ultrafast light-matter interaction with twist control of such band structures hence suggests a particularly promising route towards nonequilibrium functionalization, with first explorations already under way \cite{topp_tbg_2019,moireqs,rodriguez-vega_moire-floquet_2020}. These efforts should of course be complemented by more traditional avenues of materials science, such as material synthesis tailored specifically to questions relevant to nonequilibrium control. For example a systematic exploration of engineerable free energy landscapes, particularly suitable for control via photoexciation, in complex oxides is highly desirable. 

These approaches expose the need for new paradigms of materials synthesis and predictions that address the requirements for ultrafast control of nonthermal states of matter. Theory needs to turn from toy models to realistic materials, to guide not only the process of directed materials synthesis, but also to provide principles for how nonthermal pathways of ultrafast control can be obtained. To achieve this leap, inspiration can be drawn from other fields, where such an integrated approach has already been established. Namely, an increased feedback between experimental characterization, theoretical understanding, and materials synthesis is a strategy that is successful employed in many branches of quantum materials research, for instance in the field of oxides \cite{OxidesRoadmap}, or in the study of quantum materials with angle-resolved photoemission spectroscopy \cite{sobota_electronic_2020}. 
This interdisciplinary approach requires additional effort, as a common language and concepts need to be developed for thinking about the underlying physics. Parts of such a program have been laid out in this Colloquium. This is but one step along the long path of unifying themes and languages with the goal of fostering the relatively young field of ultrafast quantum materials science. 

\section*{Acknowledgments} 

We are grateful to H.~Aoki, R.~Averitt, M.~Bonitz, A.~Cavalleri, M.~Dean, P.~Hofmann, D.~Juraschek, P.~Kirchmann, A.~Kogar,  \mbox{W.-S}.~Lee, D.~Mazzone, M.~Mitrano, P.~Narang, D.~Nicoletti, B.~Normand, H.~Petek, J.~Ravnik, G. Refael, R.~Tuovinen and A.~Zong for stimulating discussions and critical feedback. DMK, JWM and MAS acknowledge support from the Max Planck-New York City Center for Non-Equilibrium Quantum Phenomena. DMK acknowledges the Deutsche Forschungsgemeinschaft (DFG, German Research Foundation) for support through RTG 1995 and under Germany's Excellence Strategy - Cluster of Excellence Matter and Light for Quantum Computing (ML4Q) EXC 2004/1 - 390534769. MC acknowledges support from a startup grant from the University of Pennsylvania, and from the Flatiron Institute, a division of the Simons Foundation. JWM acknowledges support from the Cluster of Excellence ‘CUI: Advanced Imaging of Matter’ of the Deutsche Forschungsgemeinschaft (DFG), EXC 2056, project ID 390715994 and is funded by the Deutsche Forschungsgemeinschaft (DFG, German Research Foundation) – SFB-925 – project 170620586. MAS acknowledges financial support through the Deutsche Forschungsgemeinschaft (DFG, German Research Foundation) via the Emmy Noether program (SE 2558/2). 

All authors contributed equally to this work.

\appendix
\section{Technical advances shaping ultrafast quantum materials science}
\label{sec:appendix}
%Technical advances, both in ultrafast pump-probe experiments and in nonequilibrium quantum many-body theory, have enabled a deeper understanding of light induced phenomena in quantum materials in recent years. We summarize some of these developments here.   
  
\subsection{Experimental tools}
\label{sec:exp_advances}

{\it Time-resolved optical spectroscopy} remains the most commonly used approach to access the time-dependent optical properties of quantum materials after photoexcitation. In recent years, the field has seen an evolution from the early measurements of transient absorption and reflectivity \cite{Schoenlein1987,Elsayed1987,Koshihara90,miyano_manganite_1997,Chemla2437,Tsen2001} into a multifaceted set of techniques, where the particular details of experiments depend on the targeted subsystem and dynamics \cite{Averitt2002,orenstein_ultrafast_2012}. Some of the most advantageous capabilities of these approaches are: (i) direct detection of transient changes in the electronic joint density of states via frequency-resolved measurements of the transient complex optical conductivity from the THz to the extreme ultraviolet range \cite{GedikStark1,Jager9558,Siegrist2019,Baldini2020}, (ii) simultaneous measurements of the dynamics of different subsystems by combining multiple detection schemes  \cite{von_hoegen_probing_2018}, including transient non-linear optical processes \cite{Sala2016,Woerner2013,Mahmood2020} and polarization rotations sensitive to changes in magnetic orders \cite{KIMEL20201,Beaurepaire96,Kirilyuk2010,Walowski2016,Nemec2018,Schlauderer2019}, (iii) the integrability with other external stimuli, for example magnetic fields and hydrostatic pressure \cite{Nicoletti_Field,Trigo_pressure,Mitrano_pressure,Cantaluppi2018}, and (iv) the ability to modulate and control the optical pulse to gain real-space information \cite{Gedik2003,Torchinsky2014,Mahmood2018}. These techniques have enabled the observation of a wide range of phenomena in the time domain including quasiparticle relaxation dynamics, electron-boson coupling strengths, gap magnitudes, photoexcited order parameters and collective mode oscillations \cite{Brorson90,Segre02,Demsar99,Demsar2002}, phase coexistence,  light-induced phase transitions \cite{Koshihara90,rini_control_2007,giannetti_ultrafast_2016,Yusupov2010}, magnetic moment precession, \cite{nova_effective_2017,afanasiev2019lightdriven,disa_polarizing_2020,stupakiewicz_ultrafast_2021}, and the relaxation dynamics of superconductors, and light-induced superconducting-like states \cite{Yu91,Demsar99_ii,Kaindl00,fausti_light-induced_2011,Hu2014,nicoletti_optically_2014,mitrano_possible_2016,cremin_photoenhanced_2019}. 

{\it Time-resolved scattering techniques.} Ultrafast X-ray scattering techniques take advantage of photon wavelengths of comparable magnitude to the atomic spacing to probe structural dynamics \cite{Fritz07,Cavalleri06,Johnson08,mankowsky_nonlinear_2014,Rettig15,Gerber15,Gerber17} or the evolution of spin, charge and orbital electronic orders \cite{Johnson12,Beaud14,lee_nickelate_2012,Kubacka14,Mitrano19,Forst11,Dean2016} after optical excitation. The way towards sub-picosecond X-ray pulses has been paved by laser-driven \cite{Rischel97} and hybrid laser-accelerator-based sources \cite{Schoenlein96,Schoenlein2000}, which in addition to diffraction \cite{Cavalleri06,Johnson08} have also enabled spectroscopic measurements \cite{Cavalleri05,Stamm07}. The advent of X-ray free-electron lasers (FEL) \cite{Emma10,Seddon17} has led to another burst of quantum materials studies, benefiting from femtosecond pulses of Ångstrom wavelength and increased brilliance. We refer to in-depth reviews on ultrafast X-ray scattering \cite{buzzi_probing_2018} and resonant inelastic X-ray scattering \cite{Cao_rev_19,Mitrano2020}. 

The wavelength tunability of pulsed X-ray sources is key to accessing different phenomena. For example, the `soft' X-ray regime ($250 - 2000$~eV) includes many strongly resonant elemental absorption edges, e.g., 3$d$~transition metals, which can be targeted to probe electronic properties. Although such resonances also exist in the `hard' X-ray regime ($>5000$~eV), its main use is to access structural properties and large portions of reciprocal space. X-ray FELs can also be used to access inelastic scattering channels in the time domain, for example to map out energy- and momentum-resolved nonequilibrium phonon dispersions \cite{Trigo13,Zhu15} and dynamics of magnetic correlations \cite{Dean2016,Mazzone20}, as well as disordering during ultrafast phase transitions \cite{Wall18}. The coherence of FEL light has also been used to directly probe fluctuating topological order via X-ray photon correlation spectroscopy \cite{Seaberg17}.

Ultrafast electron diffraction (UED) is a complementary scattering technique that also directly measures ultrafast structural dynamics in solids \cite{Siwick03,Zewail2006,baum_attosecond_2007}. The larger scattering cross section of electrons, compared to X-ray photons, makes UED particularly suited to studying thin samples. New MeV electron sources \cite{Weathersby2015}, THz streaking \cite{Kealhofer16,Zhao18} and pulse compression schemes \cite{Qi20,Kim20} can now achieve $<100$~fs time resolution. Recent experiments have taken advantage of these capabilities, e.g., to investigate lattice dynamics \cite{Mannebach15,Waldecker2017,Konstantinova2018} and electronic orders \cite{Vogelgesang2018,Guyader17,Zong18,Sie2019,Haupt16}, as well as to study electron-phonon couplings \cite{Waldecker2016,Harb2016,Stern2018,Cotret2019,Horstmann2020}.

\textit{Time- and angle-resolved photoemission spectroscopy} (tr-ARPES) measures changes in the band structure and single-particle spectral function of solids with momentum resolution \cite{Haight1988,petek97,Bovensiepen12,Smallwood2016,Gedik2017,Lv2019,Nicholson821,Wegkamp14,Zhou2018}. This technique has been used, for example, to study decoherence effects in the excitation process \cite{Ogawa1997,Hofer1480,reutzelPRX}, to shed light on the physics of high-temperature superconductors \cite{Avigo13,Smallwood12,Parham2017,Yang19,Yang14}, to track the melting and recovery of charge-density wave orders \cite{Rohwer2011,Perfetti06,Hellmann2010,schmitt_transient_2008,Hellmann2012,Ligges18,Zong2019,Rettig2016}, to directly probe excitonic states \cite{Madeo1199,Cui2014}, to measure the relaxation dynamics of photocurrents \cite{Subcycle3,GuddeReview} and the coupling between electronic and lattice degrees of freedom \cite{Kemper2017,Gerber17,Na2019}. tr-ARPES also permits the detection of transiently populated topological states \cite{Sobota12,Sobota13,Belopolski2017,Zhang2017}, observation of Floquet-Bloch states \cite{GedikARPES1,GedikARPES2}, and identification of nonthermal electronic regimes \cite{HoffmanHotGraphene,Gierz2013,Nonthermal_Graphene_Damascelli_2020}. An exciting prospect is the implementation of new detection schemes extending time- and momentum-resolved microscopy to FELs \cite{ARPES_FEL}.

\textit{Time-resolved scanning probes}. Time-resolved scanning near-field optical microscopy (tr-SNOM) tracks photoinduced changes in the optical constants of  materials on the 10~nm length scale with high temporal and spectral resolution \cite{snomEisele,snomWagner}. tr-SNOM has been used, for example, to investigate photoinduced insulator-to-metal transitions \cite{snomDonges, snomHuber1} and image the propagation of plasmon-polaritons in a variety of systems \cite{snomWagner, snomBasov,snomHuber2}. %A multi-messenger approach using SNOM, AFM, and magnetic force microscopy has also been used to study and manipulate photoinduced metastable metallic states in strained films \cite{snomMcLeod}. 
Ultrafast scanning tunneling microscopy (STM) is also gaining traction and probes quantum tunneling at the atomic length scale with sub-femtosecond time resolution \cite{stmNunes, stmCocker1,stmKern}. Ultrafast STM has been used to track carrier and spin dynamics in photoexcited semiconductors  \cite{stmShigekawa1,stmShigekawa2}, the ultrafast vibrational motion of a single molecule \cite{stmCocker2}, and image surface plasmons in gold \cite{stmKern}. %Photoinduced metastable amorphous states were also recently investigated using STM \cite{stmMihailovic}. 

\textit{Time-resolved transport}. 
Microstructured devices incorporating laser-triggered photoconductive switches \cite{Auston} have been used to investigate a variety of ultrafast transport phenomena, such as ballistic electron flow in carbon nanotubes \cite{McEuen}, helicity-dependent photocurrents in topological insulators \cite{Holleitner}, and the transport properties of Floquet-Bloch states in graphene \cite{McIver}. Ultrafast transport dynamics have also been probed in carbon nanotubes \cite{Gabor2012}, graphene \cite{Xu2012}, and various van der Waals heterostructures \cite{Koppens2016,Pablo2016,Gabor2019} by measuring the photocurrent generated in response to two time-delayed laser pulses. %Beyond providing access to electrical transport, microstructured devices offer the ability to tune a material’s carrier density, apply displacement fields, run electrical currents, and study samples that are too small to be accessed using far-field optical techniques. 

\subsection{Theoretical tools}

%Most statistical and computational techniques available in thermodynamic equilibrium cannot be straightforwardly generalized to systems driven far from equilibrium. This makes the theoretical description of the out-of-equilibrium microscopic dynamics of a quantum many-body system particularly challenging

\label{sec:theory_advances}

{\it Time-dependent density functional theory} (TDDFT) is an extension of ground-state density functional theory, with similar merits and challenges \cite{marques_fundamentals_2012,ullrich_time-dependent_2012}. The full time-dependent Schr\"odinger equation for the many-particle wavefunction is replaced by an auxiliary set of Schr\"odinger (Kohn-Sham) equations, which determine the exact time-dependent density of the system \cite{Runge_Gross_1984}. Coulomb interactions are accounted for via exchange-correlation potentials, though their exact forms are unknown in practice. TDDFT has been successfully applied to a host of systems in weak and strong external driving fields, ranging from atoms and molecules to periodic solids \cite{giovannini_simulating_2013,Lian2020}. The strength of TDDFT lies in the ability to describe full-fledged material-specific details. However, the faithful description of nontrivial correlation effects still poses a severe challenge for TDDFT-based methods, in particular in pump-probe settings. 

{\it Nonequilibrium Green's functions} and diagrammatic techniques provide a framework for obtaining few-body correlation functions of time-dependent problems without computing the actual many-body wave functions \cite{stefanucci_nonequilibrium_2013}. Green's function techniques are based on single-particle propagators that describe the probability amplitude for particles (electrons, phonons, etc.) to travel between two space-time points while interacting with the rest of the system. These many-body interactions are described through the many-body self-energy. Notable approximations to calculate the self-energy include the so-called GW approximation for screened interactions \cite{aryasetiawan_thegwmethod_1998,thygesen_nonequilibrium_2007}, nonequilibrium dynamical mean-field theory (DMFT) for strongly correlated systems, including Mott insulators with mainly local self-energies \cite{PhysRevB.78.235124,freericks_nonequilibrium_2006,aoki_nonequilibrium_2014}, and the real-time functional renormalization group (FRG) approach \cite{PhysRevB.85.085113}. In a nonequilibrium setup these approximation schemes are more difficult to handle than in equilibrium because time-translational invariance cannot be exploited and, in general, two time variables instead of one energy variable must be kept. Nonequilibrium Green's functions are an ideal starting point when approximations for the self-energy can be physically motivated, e.g., perturbative weak- or strong-coupling expansions, or the local-self-energy approximation of DMFT. 

Green's function-based real-time diagrammatic quantum Monte Carlo techniques provide alternative and \textit{a priori} controlled solutions via stochastic sampling of a perturbation series. While these approaches suffer from a notorious `dynamical sign problem' \cite{PhysRevB.79.035320}, whereby the non-positivity of sampling weights results in a computational effort that increases exponentially in time, recent algorithmic advances with modest computational scaling including the `inchworm' algorithm \cite{PhysRevLett.115.266802} and conformal transformations \cite{PhysRevX.9.041008} have been put forward. Real-time diagrammatic quantum Monte Carlo is the method of choice whenever the sign problem can be averted. 

{\it Tensor networks} 
provide a faithful representation of many-body states and operators as a `contraction' network of tensors that encode the local properties of the system \cite{Cirac08,Schollwock11}. In one dimension matrix product states are one particularly prominent example of a tensor network and provide the basis for elegant density matrix renormalization group implementations \cite{Schollwock11}. In general, the representation of generic quantum states requires contracting tensor dimensions that scale with the exponential size of the full many-body Hilbert space. However, in many physically relevant situations the notion of locality introduced by the tensor network allows one to represent states with low entanglement via tensors of much reduced dimensionality. In this sense, tensor networks might be viewed as a low-entanglement method. The entanglement of ground states generally scales favorably in low-dimensional systems due to the area law \cite{ORUS2014117,EisertArea}.
Conversely, dynamics far from equilibrium typically involve highly excited states with volume law scaling of the entanglement entropy. Therefore, in nonequilibrium studies only short timescales can be simulated. Numerous methods for time-evolution have been proposed and benchmarked, including the block decimation, matrix product operator techniques, Krylov methods, or the time-dependent variational principle \cite{paeckel19}. In a nutshell, tensor networks are useful whenever entanglement entropy can be kept at bay. 

{\it Other variational techniques.} Beyond asymptotically exact representations of the many-body wavefunction, simpler variational starting points serve as theoretical tools to gain insight into specific physical systems, often based on some intuition of the relevant physics. Examples include the time-dependent Gutzwiller approximation \cite{seibold01,schiro10}, variational Monte Carlo \cite{carleo11}, Gaussian and non-Gaussian variational states \cite{shi17,hackl20}, or Gross-Pitaevskii equations \cite{gross61,pitaevskii61}. More recently, machine learning-inspired restricted Boltzmann machines have emerged, which provide a more flexible variational subspace for time evolution \cite{carleo16}.
To summarize, common to all these techniques are Ans\"atze for the many-body wavefunction with an economical number of free parameters, which are governed by equations of motion as determined via the time-dependent variational principle. Though, the general usefulness of these techniques for simulations related to pump-probe experiments is not quite clear yet, they can provide a good starting point whenever a suitable variational manifold can be identified. 

\bibliography{references_DK,references_MC,references_MS,references_JM,references_ADLT,references_SG,references_ADLT_ii}
%\bibliography{ref_all}

\end{document}